\documentclass[sigconf]{acmart}

\usepackage{graphicx}
\usepackage{booktabs} 
\usepackage{subfigure}
\usepackage[]{amsmath,amsfonts,amssymb}
\usepackage{breqn} 
\usepackage{mathtools}
\usepackage{siunitx}
\usepackage{epstopdf}
\usepackage{natbib}
\usepackage{fancyhdr}
\DeclareMathOperator*{\argmax}{arg\,max}

\copyrightyear{2018} 
\acmYear{2018} 
\setcopyright{acmcopyright}
\acmConference[NANOCOM '18]{NANOCOM '18: ACM The Fifth Annual International Conference on Nanoscale Computing and Communication}{September 5--7, 2018}{Reykjavik, Iceland}
\acmBooktitle{NANOCOM '18: NANOCOM '18: ACM The Fifth Annual International Conference on Nanoscale Computing and Communication, September 5--7, 2018, Reykjavik, Iceland}
\acmPrice{15.00}
\acmDOI{10.1145/3233188.3233210}
\acmISBN{978-1-4503-5711-1/18/09}

\begin{document}
	\title{Direction of Arrival Estimation for Nanoscale Sensor Networks}
	
	\author{Shree Prasad M.}
	\affiliation{%
		\institution{Department of Electronics and Communication Engineering\\National Institute of Technology Goa}
		\city{Farmagudi}
		\state{Goa}
		\postcode{403401}
	}
	\email{shreeprasadm@gmail.com}
	
	\author{Trilochan Panigrahi}
	\affiliation{%
		\institution{Department of Electronics and Communication Engineering\\National Institute of Technology Goa}
		\city{Farmagudi}
		\state{Goa}
		\postcode{403401}
	}
	\email{tpanigrahi@nitgoa.ac.in}
	
	\author{Mahbub Hassan}
	\affiliation{%
		\institution{School of Computer Science and Engineering\\ University of New South Wales}
		\city{Sydney}
		\state{Australia}
	}
	\email{mahbub@cse.unsw.edu.au}
	\begin{abstract}
		Nanoscale wireless sensor networks (NWSNs) could be within reach soon using graphene-based antennas, which resonate in 0.1-10 terahertz band. To conserve the limited energy available at nanoscale, it is expected that NWSNs will communicate using extremely short pulses on the order of femtoseconds. Accurate estimation of direction of arrival (DOA) for such terahertz pulses will help realize many useful applications for NWSNs. In this paper, using the well-known MUltiple SIgnal Classification (MUSIC) algorithm, we study DOA estimation for NWSNs for different energy levels, distances, pulse shapes, and frequencies. Our analyses reveal that the best DOA estimation is achieved with the first order Gaussian pulses, which emit their peak energy at 6 THz. Based on Monte Carlo simulations, we demonstrate that MUSIC algorithm is capable of estimating DOA with root mean square error less than one degree from a distance of around 6 meter for pulse energy as little as 1 atto Joule.
	\end{abstract}
	%
	%
	\vspace{-10mm}
	\maketitle
	\vspace{-4mm}
	\section{Introduction}
	Significant progress in nanotechnology has made it possible to fabricate nanosensors from novel nano-materials that can sense chemical compounds in extremely low concentrations, and viruses or bacteria in very small populations. The limited sensing range of these nanosensors can be extended by enabling wireless communication between them, which creates the so called nanoscale wireless sensor networks (NWSNs). With wider coverage, NWSNs promise new sensing and detection applications in biomedical, environmental, industrial and military fields \cite{NANOCOMP}. 
	
	The nanoscale form factor imposes several interesting characteristics in NWSNs. Nano devices are expected to communicate in terahertz band (0.1 - 10 THz), as the nano antenna made of graphene is shown to efficiently radiate in this band \cite{GRAANT}. Due to the size and energy constraints of nanosensor devices, carrier-less pulse based communication scheme is being considered for NWSN. Gaussian pulse and its higher order time derivative of few hundred femtoseconds long duration is used to represent symbols for such communication \cite{FEMSEC}. Researchers are using these new properties to explore new communication, sensing, and detection methods in NWSNs.
	
	The use of localization techniques such as direction of arrival (DOA), time of arrival (TOA) and received signal strength (RSS) has many benefits for NWSNs.  For example, works that aim to detect and localize events from a single pulse transmission \cite{HASSR2,HASSR3}, can use localization techniques to localize the sensor node that detected the event and transmitted the pulse. Beamforming research in terahertz networks \cite{TBMAC} can exploit DOA to select the beam direction using in-band terahertz signals. Localization techniques in NWSN can be used to track the mobility of events at nanoscale. Many other benefits of localization can be envisaged. Among the three different localization techniques, DOA estimation technique is more suitable for NWSNs due to the following reasons \cite{UWB_BOOK}: 1) TOA technique requires accurate timing synchronization between transmitter and receiver. Hence it is difficult to have extra circuity for achieving timing synchronization in a size constrained nanosensor node. 2) RSS technique requires prior knowledge of  channel characteristics and transmit power by nanosesnor node. Since the nanosensor devices are expected to harvest energy from its ambient environment, the a priori knowledge of transmitted power is not known. 
	
	Although DOA is an established field of research with numerous works reported for many different scenarios, to the best of our knowledge, its potential for NWSNs has not been explored in the literature. The closest work reported is DOA estimation in pulse-based ultra-wideband (UWB) networks \cite{UWB_DOA1,UWB_DOA2}. However, these works are restricted to a few GHz bandwidth with nanosecond pulses emitting relatively high energy as such devices are expected to operate with large batteries or even grid power. NWSNs use frequencies in the THz band, extremely short pulses on the order of femtoseconds, which contains signal within terahertz bandwidth. The channel propagation in the terahertz band is affected by molecular noise, which introduces further dynamics to DOA. 
	
	The aim of this paper is to systematically study the potentials of DOA in NWSNs by varying a range of parameters for the terahertz Gaussian pulse including energy, order of time derivative, center frequency, and bandwidth. We also study the accuracy of DOA estimation as a function of distance between the emitter and the sensor array at the detection point. Our contributions can be summarized as follows:
	\begin{itemize}
		\item Using the terahertz channel characteristics for NWSN, we compare the DOA estimation performance of different higher time derivative order Gaussian pulse with different center frequencies using MUSIC algorithm. The comparisons are obtained by varying the pulse energy as well as the distance between the emitter and the array receiver. We also study the impact of the number of snapshots on the performance of DOA estimates. To the best of our knowledge, this is the first such study. We obtain several important and interesting results.
		\item Our analyses show that DOA performance improves with lower orders and higher frequencies. This suggests that for NWSNs, the best performance for DOA can be achieved with the first order Gaussian pulses emitting their peak energies at 6 THz center frequency. \textit{We find that 
			MUSIC can achieve DOA with RMSE less than one degree from a distance of 6\textit{meters} for pulse energy as little as 1 atto Joule.} This is a remarkable performance for MUSIC, which can enable many applications in NWSNs. 
		\item Our study reveals that for high frequency, distance has no significant impact for up to 1 cm, but RMSE starts to increase rapidly after that. In contrast, for low frequency, RMSE increases for increasing distance even below 1 cm. 
		\item For high frequency, pulse energy has little effect on DOA performance. Similarly, pulse energy has little impact on DOA estimate for high order and low frequency. Pulse energy, however, has significant effect on DOA performance for low order and low frequency pulses, in which case RMSE increases rapidly with decreasing pulse energy.
		
	\end{itemize}
	The rest of the paper is structured as follows. Terahertz channel impulse response and molecular absorption noise are reviewed in Section 2.  In section 3, system model is described and the application of wideband MUSIC algorithm is considered for DOA estimation of higher order Gaussian pulses for terahertz channel. In section 4, the performance of MUSIC DOA estimation algorithm is evaluated for different higher order Gaussian pulses with different center frequencies. Section 5 concludes the paper.
	\vspace{-2.5mm}
	\section{Terahertz Channel}
	The propagation of electromagnetic waves in Terahertz channel depends on its chemical composition as it is affected due to attenuation and molecular absorption noise.  The chemical composition of the terahertz channel is characterized by medium absorption coefficient $k\left(f \right)$ at frequency $f$. The effect of channel attenuation and molecular absorption noise will be significant at molecular resonant frequencies for long range communication distances. In the following subsections, the characteristics of terahertz channel and molecular absorption noise is described\cite{FEMSEC}.
	\vspace{-2.5mm}
	\subsection{Terahertz Channel Impulse Response}
	The pulse propagating in a terahertz channel suffers attenuation due to the spreading loss and the molecular absorption loss. The terahertz channel impulse response accounting for both the spreading loss and the molecular absorption loss in frequency domain is represented as\cite{FEMSEC}  
	\begin{equation}\label{eq:chresp}
	H\left( f,d_{r}\right) = H_{spread}\left( f,d_{r}\right) H_{abs}\left( f,d_{r}\right)  
	\end{equation}
	where $H_{spread}\left(f,d_{r} \right) $ and $H_{abs}\left(f,d_{r} \right) $ represents spreading loss and molecular absorption loss, and are given by
	\begin{equation}
	H_{spread}\left( f,d_{r}\right) =\left(  \frac{c_{o}}{4\pi d_{r} f_{o}}\right) \exp\left( -\frac{j 2\pi f d_{r}}{c_{o}}\right)  
	\end{equation}
	\begin{equation}
	H_{abs}\left( f,d_{r}\right) = \exp(-0.5 k\left(f \right)d_{r} ) 
	\end{equation}
	where $f$ denotes frequency, $c_{o}$ is the velocity of light in vacuum, $f_{o}$ is the center frequency of the graphene antenna, $d_{r}$ is the path length, and $k\left( f\right) $ is the medium absorption coefficient. The medium absorption coefficient $k\left( f\right) $ of the terahertz channel at frequency $f$ composed of $Q$ type molecules is given as
	\begin{equation}
	k\left( f\right)  = \sum_{q=1}^{Q}x_{q}K_{q}\left( f\right).
	\end{equation} 
	where $x_{q}$ is the mole fraction of molecule type $q$ and $K_{q}$ is the absorption coefficient of individual molecular species.
	\vspace{-2.5mm}
	\subsection{Molecular Absorption Noise}
	In terahertz channel, the main contribution for ambient noise comes from the molecular absorption noise. This noise arises because of re-radiation of part of wave energy that was absorbed by the molecules in the medium during transmission of the pulses. Hence, molecular absorption noise is present at the receiver only during transmission of pulses. The total molecular absorption noise p.s.d. $S_{N}\left( f,d_{r}\right) $  is the sum of background atmospheric noise p.s.d \(S_{N_{B}}\left( f,d_{r}\right)\) and the self induced noise p.s.d. $S_{N_{P}}\left(f,d_{r} \right)$ and is given as\cite{FEMSEC} 
	\begin{equation}
	S_{N}\left( f,d_{r}\right)  = S_{N_{B}}\left(f,d_{r} \right)+S_{N_{P}}\left(f,d_{r} \right)  
	\end{equation}
	\begin{equation}\label{eqn:mnm1}
	S_{N_{B}}(f, d_{r}) = \lim\limits_{d_{r} \rightarrow \infty} k_{B} T_{0}\left( 1-\exp\left( -k\left(f \right)d_{r} \right) \right) \left( \frac{c_{0}}{\sqrt{4\pi}f_{0}}\right)^{2} 
	\end{equation}
	\begin{equation}\label{eqn:mnm2}
	S_{N_{P}}\left(f,d_{r} \right) = S_{P}\left( f\right)\left( 1-\exp\left( -k\left(f \right)d_{r} \right) \right) \left( \frac{c_{0}}{4\pi d_{r} f_{0}}\right)^{2} 
	\end{equation}
	where $k_{B}$ is the Boltzmann constant, $T_{0}$ is the room temperature and $S_{P}\left( f\right) $ represents p.s.d. of transmitted pulse. 
	\vspace{-2mm}
	\begin{figure}[H]
		\centering
		\includegraphics[width=0.9\columnwidth, height = 4.5cm]{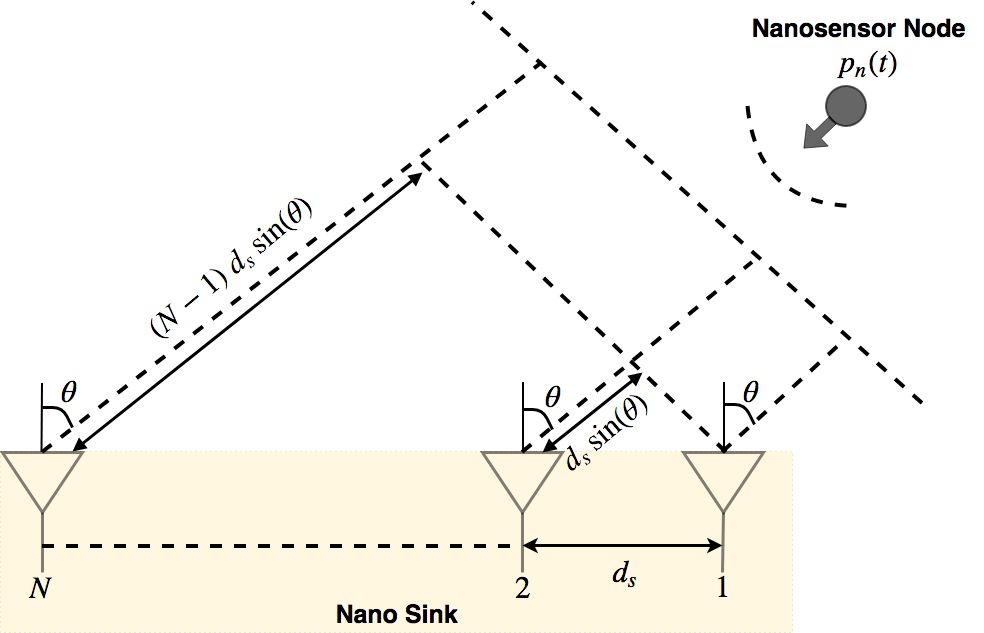}
		\vspace{-1mm}
		\caption{N-element ULA at the nanosink receiving pulses from nearby nanosensors}
		\label{fig:DOA_E}
		\vspace{-4mm}
	\end{figure}
	\section{System Model}
	The DOA estimation of nanosensor nodes is performed by a nano sink as shown in Fig. \ref{fig:DOA_E}. The nano sink consists of uniform linear array (ULA) with $N$ antenna elements. Nanosensor nodes in NWSN is envisioned to communicate using carrier-less pulse based transmission. In this paper, these pulses are assumed as higher order time derivatives of Gaussian pulses of few hundred femtoseconds duration. The Fourier representation of time derivative of \text{the} Gaussian pulse with order of derivative $n$ is also Gaussian shaped and is represented as 
	\vspace{-1mm}
	\begin{equation}
	G_{n}\left( f\right)  = \left(j 2\pi f\right)^{n} a_{n} e^{-0.5\left( 2\pi\sigma f\right)^{2} }
	\end{equation}
	where $a_{n}$ is the normalizing constant to adjust the pulse energy, and $\sigma$ is the standard deviation of the Gaussian pulse in seconds. The center frequency of p.s.d. of Gaussian pulse increases with time derivative order and is represented as \cite{FMFRE}
	\begin{equation}
	f_{c} = \frac{\sqrt{n}}{2\pi \sigma}
	\end{equation} 
	The higher order Gaussian pulses transmitted by nanosensor nodes is assumed to be uniformly spaced in time and is represented as
	\vspace{-1.5mm}
	\begin{equation}
	p_{n}\left( t\right)  = \sum_{r=-\infty}^{\infty}a_{r}g_{n}\left( t-rT_{p}\right) 
	\vspace{-1mm}
	\end{equation}
	where $g_{n}\left( t\right) $ represents Gaussian pulse of order $n$, $a_{r}$ is the information symbol sequence and $T_{p}$ is defined as total pulse duration of Gaussian pulse. Though the higher order Gaussian pulses have infinite time duration, here it is assumed that $T_{p} = 10\sigma$ \cite{FMFRE}. $T_{p}$ is the time interval which contains more than $99.99\%$ of pulse energy. Assuming Bi-Phase modulation is used to represent information symbol\cite{FMFRE,BPSK_HAS_SIR}, the PSD of transmitted signal $p_{n}\left( t\right)$  by nanosesnor device is represented as
	\vspace{-0.75mm}
	\begin{equation}
	\resizebox{0.25\columnwidth}{!}{$S_{n}\left(f \right)  = \frac{\left| G_{n}\left( f\right) \right|^{2} }{T_{p}}$}
	\vspace{-1mm}
	\end{equation}
	\vspace{-5mm}
	\subsection{Problem Formulation}
	\label{sec:PF}
	Consider a ULA with inter-element spacing of $d_{s}\:\text{m }$ as shown in  Fig.\ref{fig:DOA_E}. Here, it is considered that a single nanosensor node to be localized is present at a distance of $d_{r}$ from ULA. This transmission distance is assumed to be in the far-field region of ULA. The received wideband higher order Gaussian pulse at the output of the $i$th element in ULA is represented as \cite{FEMSEC}. 
	\vspace{-0.5mm}
	\begin{equation}\label{eq:datam}
	y_{i}\left(t, d_{r}\right) =  p_{n}\left(t - \tau_{i} \right) * h\left(t,d_{r} \right) + v_{i}\left( t,d_{r}\right)   
	\end{equation}
	\begin{equation}\label{eq:td}
	\tau_{i} = \left[ \left(i-1\right) d_{s}\sin\left( \theta\right)\right]  /c_{o} 
	\end{equation}
	where $h\left(t,d_{r}\right) $ is the terahertz channel impulse response between ULA and nanosensor node in $\theta$ direction. $\tau_{i}$ represents time delay at $i^{\text{th}}$ antenna element of ULA with reference to antenna element $1$. $v_{i}$ represents molecular absorption noise created between element $i$ of ULA and nanosensor node. When the signal received at the output of the ULA is observed for sufficiently large time interval $\Delta T$, the Fourier representation of  \eqref{eq:datam} is given as\cite{OPVT}
	\vspace{-0.5mm}
	\begin{equation}\label{eq:datafm}
	\resizebox{0.95\columnwidth}{!}{$Y_{i}\left( f_{l},d_{r}\right) =  e^{-j2\pi f_{l} \tau_{i}} P_{n}\left( f_{l}\right) H\left( f_{l},d_{r}\right) + V_{i,}\left( f_{l},d_{r}\right) , \: \text{for}\;\; l = 0,\cdots, L$}
	\end{equation}
	where $f_{l} $ is the frequency bin, $P_{n}\left( f_{l}\right) $, $H\left( f_{l},d_{r}\right) $, and $V_{i,}\left( f,d_{r}\right)$ are Fourier coefficients of Gaussian pulse, terahertz channel impulse response and molecular absorption noise respectively. Further the output of array is observed for $K$ non-overlapping time interval $\Delta T$ and Fourier coefficients is computed for each time interval. Here, $K$ is called as frequency snapshot number. The number of frequency bins $L$ is given as 
	\vspace{-1mm}
	\begin{equation}\label{eq:timbw}
	L = \lfloor B \cdot \Delta T \rfloor +1
	\end{equation}
	where $\lfloor \cdot  \rfloor$ is the floor operator, $B$ is the terahertz channel bandwidth and the observation time interval $\Delta T$ is significantly greater than the propagation time across ULA. For a  Gaussian pulse of particular order and center frequency, $\frac{\Delta T}{T_{p}}$ number of pulses are received at ULA over single observation interval $\Delta T$.
	Now, the Fourier coefficients at frequency $f_{l}$ across $N$ sensors for $K$ number of frequency snapshots is represented in matrix form as 
	\vspace{-1mm}
	\begin{equation}\label{eq:datafm_vec}
	\boldsymbol{Y}\left(f_{l},d_{r} \right) =  H\left( f_{l},d_{r}\right) \boldsymbol{a}\left(f_{l} \right) \boldsymbol{P}_{n}\left(f_{l} \right) + \boldsymbol{V}\left( f_{l},d_{r}\right) 
	\end{equation}
	where $\boldsymbol{Y}\left(f_{l},d_{r} \right) \in \mathbb{C}^{N \times K}$, $\boldsymbol{V}\left( f_{l},d_{r}\right)\in \mathbb{C}^{N \times K}$ and\\ \resizebox{0.47\columnwidth}{!}{$\boldsymbol{P}_{n}\left(f_{l} \right)\overset{\Delta}{=}\left[ {P}_{n1}\left(f_{l} \right),\cdots,{P}_{nK}\left(f_{l} \right)\right] $}. \resizebox{0.47\columnwidth}{!}{$\boldsymbol{a}\left(f_{l} \right) =\left[1, e^{-j2\pi f_{l} \tau_{1}}, \cdots, e^{-j2\pi f_{l} \tau_{N}}\right]^{T} $} is the array manifold vector. 
	The covariance matrix $\boldsymbol{R_{Y}}\left(f_{l},d_{r} \right)$ of $\boldsymbol{Y}\left(f_{l},d_{r} \right)$ is given as 
	\vspace{-1mm}
	\begin{equation}\label{eq:covmat}
	\boldsymbol{R_{Y}}\left(f_{l},d_{r} \right)  = \mathbb{E}\left[ \boldsymbol{Y}\left(f_{l},d_{r} \right)\boldsymbol{Y}\left(f_{l},d_{r} \right)^{H}\right] 
	\end{equation}
	where $\left(\cdot \right) ^{H}$ denotes conjugate transpose and $\mathbb{E}\left[ \cdot\right] $ represents expectation.
	The term $\left[ \boldsymbol{Y}\left(f_{l},d_{r} \right)\boldsymbol{Y}\left(f_{l},d_{r} \right)^{H}\right] $ in \eqref{eq:covmat} is 
	\vspace{-1.5mm}
	\begin{dmath}\label{eq:YYHEXP}
		\resizebox{0.9\columnwidth}{!}{$H\left( f_{l},d_{r}\right) \boldsymbol{a}\left( f_{l}\right) \boldsymbol{P}_{n}\left( f_{l}\right)\cdot  \left(  H\left( f_{l},d_{r}\right) \boldsymbol{a}\left( f_{l}\right) \boldsymbol{P}_{n}\left( f_{l}\right)\right)^{H}+\boldsymbol{V}\left( f_{l},d_{r}\right)\boldsymbol{V}^{H}\left( f_{l},d_{r}\right)$} 
		+
		\resizebox{0.9\columnwidth}{!}{$H\left( f_{l},d_{r}\right) \boldsymbol{a}\left( f_{l}\right) \boldsymbol{P}_{n}\left( f_{l}\right)\cdot  \left( \boldsymbol{V}\left(f_{l},d_{r} \right)  \right)^{H}+ \boldsymbol{V}\left(f_{l},d_{r}\right)  \left(  H\left( f_{l},d_{r}\right) \boldsymbol{a}\left( f_{l}\right) \boldsymbol{P}_{n}\left( f_{l}\right)\right)^{H}$}
	\end{dmath}
	\vspace{-2.70mm}
	Taking expectation on \eqref{eq:YYHEXP} and since self induced noise is correlated with the transmitted pulse and using \eqref{eq:chresp} and \eqref{eqn:mnm2}, the expectation of third term on the right hand side of \eqref{eq:YYHEXP} reduces to 
	\begin{dmath}\label{eq:CORSIGNOI}
		\resizebox{0.3\textwidth}{!}{$\mathbb{E}\left[ \left(  H\left( f_{l},d_{r}\right) \boldsymbol{a}\left( f_{l}\right) \boldsymbol{P}_{n}\left( f_{l}\right)\right)\cdot  \left( \boldsymbol{V}\left(f_{l},d_{r} \right)  \right)^{H}\right]$} \\
		\resizebox{0.425\textwidth}{!}{$= \left| \left|\boldsymbol{P}_{n} \right|\right|  _{2}\left( \frac{c_{0}}{4\pi d_{r} f_{o}}\right) ^{2}\frac{\sqrt{\left(1-\exp\left(-x  \right)  \right) }}{\exp\left(0.5x\right)}  \exp\left(-j\frac{2\pi f d_{r}}{c_{o}} \right) \boldsymbol{a}_{m}\left( f_{l}\right)  \boldsymbol{1}_{1\times N}$}
	\end{dmath}
	where $\left| \left|\cdot \right| \right|_{2} $ denotes $l_{2}$ norm,  $x = k\left( f_{l}\right)\cdot d_{r}$  and $\boldsymbol{1}_{1\times N}$ is ones vector of size $1\times N$. The term $\left(\frac{c_{o}}{4\pi d_{r} f_{o}} \right)^{2} \ll 1 $, as center frequency $f_{c}$ of higher order Gaussian pulses is considered to be greater than 2 THz. Based on these assumptions \eqref{eq:CORSIGNOI} is approximated as zero. Similar arguments can be made for fourth term in \eqref{eq:YYHEXP} and can approximated to zero.
	Based on the above assumptions, \eqref{eq:covmat} is simplified as 
	\vspace{-1.5mm}
	\begin{equation}\label{eq:stdeqn}
	\resizebox{0.425\textwidth}{!}{$\boldsymbol{R_{Y}}\left(f_{l},d_{r} \right)   =	\left|P_{n}\left( f_{l}\right) \right|^{2} \left|H\left( f_{l},d_{r}\right) \right|^{2} \boldsymbol{a}\left( f_{l}\right)\boldsymbol{a}\left( f_{l}\right)^{H} +\sigma^{2}\left( f_{l},d_{r}\right)  \boldsymbol{I}_{N}$}
	\end{equation}
	here $\mathbb{E}\left[ \boldsymbol{V}\left( f_{l},d_{r}\right)\boldsymbol{V}^{H}\left( f_{l},d_{r}\right) \right] =\sigma^{2}\left( f_{l},d_{r}\right)$  is the noise variance due to transmission of Gaussian pulse around narrow frequency sub-band centered at frequency $f_{l}$ and $\boldsymbol{I}_{N}$ is the identity matrix of size $N\times N$. Eqn.\eqref{eq:stdeqn} is same as the covariance matrix at the output of ULA assuming noise to be independent of Gaussian pulses emitted by nanosensor devices. $\sigma^{2}\left( f_{l},d_{r}\right)$ is computed as 
	\begin{equation}
	\resizebox{0.2\textwidth}{!}{$\sigma^{2}\left( f_{l},d_{r}\right) = \int S_{N}(f_{l},d_{r}) df$}
	\end{equation}
	\vspace{-5mm}
	\subsection{DOA estimation of Gaussian Pulses}
	Wideband DOA estimation methods is used for localizing nano-devices due to ultra wide frequency bandwidth of higher order Gaussian pulses. Here, incoherent multiple signal classification (IMUSIC) DOA estimation technique is used, which is based on eigen value decomposition (EVD) of the received covariance matrix $\boldsymbol{R_{Y}}\left(f_{l},d_{r} \right)$ and it is represented as
	\vspace{-1.5mm}
	\begin{dmath}
		\boldsymbol{R_{Y}}\left(f_{l},d_{r} \right) = \boldsymbol{E}_{s}\left(f_{l},d_{r} \right)\boldsymbol{\Lambda}_{s}\left(f_{l},d_{r} \right)\boldsymbol{E}_{s}^{H}\left(f_{l},d_{r} \right) \\
		+\boldsymbol{E}_{n}\left(f_{l},d_{r} \right)\boldsymbol{\Lambda}_{n}\left(f_{l},d_{r} \right)\boldsymbol{E}_{n}^{H}\left(f_{l},d_{r} \right)
	\end{dmath}
	where $\boldsymbol{E}_{s}\left(f_{l},d_{r} \right)$ and $\boldsymbol{E}_{n}\left(f_{l},d_{r} \right)$ are signal and noise eigenvector matrix and $\boldsymbol{\Lambda}_{s}\left(f_{l},d_{r} \right)$ and $\boldsymbol{\Lambda}_{n}\left(f_{l},d_{r} \right)$ are diagonal matrix corresponding to the eigenvalues of signal and noise vector space.\\
	In IMUSIC DOA estimation technique, narrowband MUSIC DOA estimation technique is independently applied to each $L$ number of frequency bins. The IMUSIC wideband DOA estimation technique is given as 
	\vspace{-1.5mm}
	\begin{equation}\label{eq:IMUS}
	P_{\text{IMUSIC}}( \hat{\boldsymbol{\theta}}, d)  = \sum_{l=1}^{L}\frac{\boldsymbol{a}^{H}\left(f_{l},\theta \right)\boldsymbol{a}\left(f_{l},\theta \right) }{\boldsymbol{a}^{H}\left(f_{l},\theta \right)\boldsymbol{E}_{n}\left(f_{l},d_{r} \right)\boldsymbol{E}_{n}^{H}\left(f_{l},d_{r} \right)\boldsymbol{a}\left(f_{l},\theta \right)}
	\end{equation}
	Eqn. \eqref{eq:IMUS} is called as IMUSIC spectrum and it is observed that, the quality of DOA estimate depends on communication distance between nanosensor device and ULA. The DOA estimate from IMUSIC spectrum is estimated as 
	\vspace{-1.5mm}
	\begin{equation}\label{eq:tht_est}
	\hat{\boldsymbol{\theta}} \left( d\right) = \argmax_{\boldsymbol{\theta}}\left[ P_{\text{IMUSIC}}( \hat{\boldsymbol{\theta}}, d)\right] 
	\end{equation}
	Further, the received covariance matrix at each frequency bin $f_{l}$ is estimated as
	\begin{equation}
	\boldsymbol{\hat{R}_{Y}}\left(f_{l},d_{r} \right) = \frac{1}{K}\boldsymbol{Y}\left(f_{l},d_{r} \right) \boldsymbol{Y}^{H}\left(f_{l},d_{r} \right)
	\end{equation}
	\section{Simulation Results}
	In this section, simulation results are presented for wideband DOA estimation of a single nanosensor device source. IMUSIC algorithm as explained in section \ref{sec:PF} is implemented using MATLAB R2014a
	\vspace{-2mm}
	\subsection{Parameters and Performance Metrics} The terahertz channel frequency band is considered from 1 THz to 10 THz. In the simulation, terahertz channel is assumed as standard air medium in summer with $1.86 \%$ concentration of water vapor molecules.  The High-resolution transmission molecular absorption (HITRAN) database is used to obtain the molecular absorption coefficient $k\left( f\right) $ of the terahertz channel\cite{HTRAN}. The accuracy of DOA estimation algorithm for single nanosensor device source is measured in terms of root mean square error (RMSE) and is defined as 
	\vspace{-1.5mm}
	\begin{equation} \label{eq:rse}
	\vspace{-0mm}
	\text{RMSE} = \sqrt{\frac{1}{N_{run}}\sum_{i=1}^{N_{run}}\left (  \hat{\theta}\left( i \right)-\theta \right )^{2}}
	\vspace{-1mm}
	\end{equation}
	where \(N_{run}\) is the number of Monte-Carlo simulations, \(\hat{\theta}\left(i\right)\) is the estimate of  DOA in the \(i^{th}\) run and \(\theta\) is the true DOA of the nanosensor device. The DOA estimate obtained in the \(i^{th}\) run corresponds to location of highest peak in the MUSIC spectrum.
	\begin{figure}[!b]
		\centering
		\subfigure[~]{
			\includegraphics[width=0.5\columnwidth, height = 3.25cm]{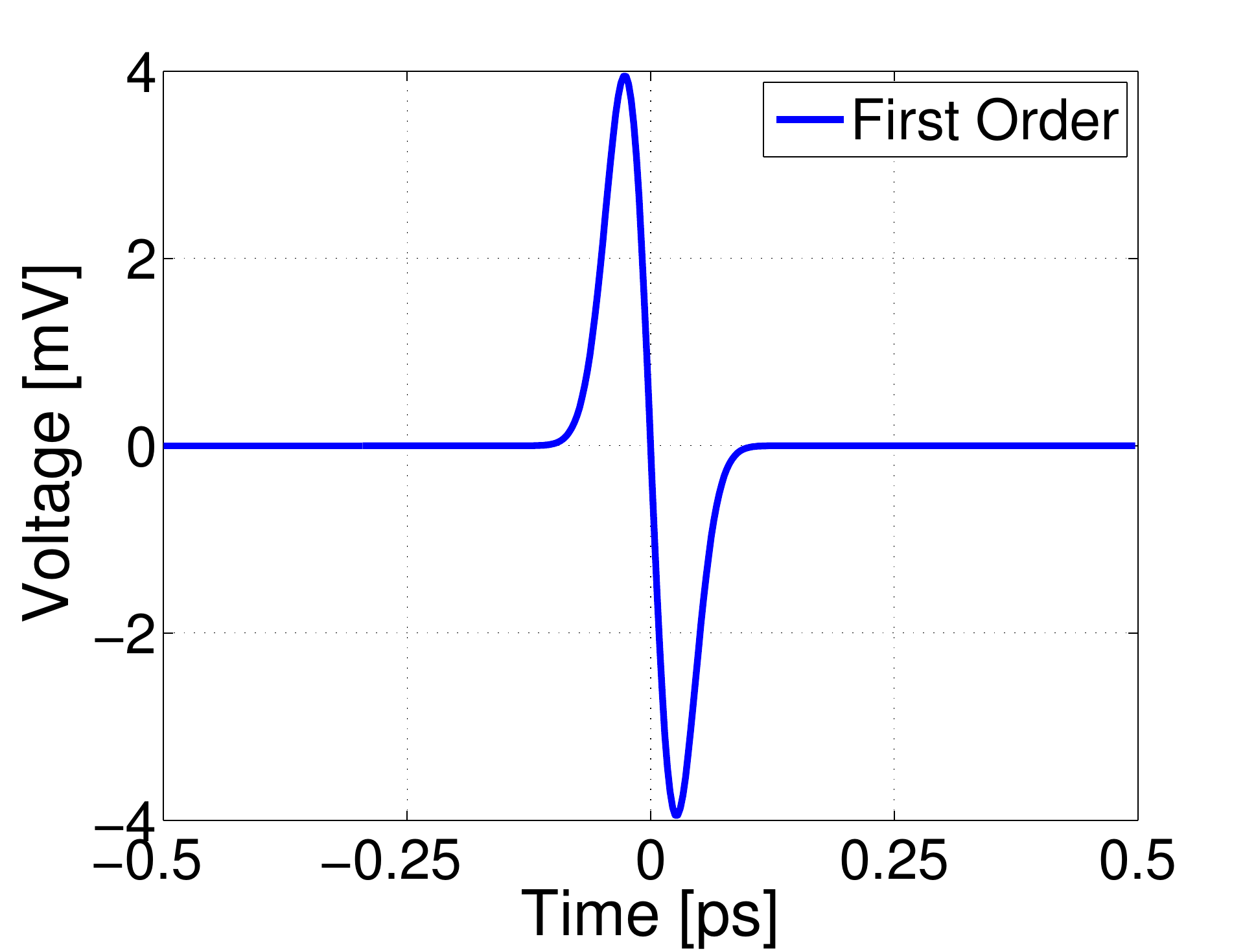}
			\label{fig:FIRSTO}
		}
	   \hspace{-5mm}
		\vspace{-4mm}
		\subfigure[~]{
			\includegraphics[width=0.5\columnwidth, height = 3.25cm]{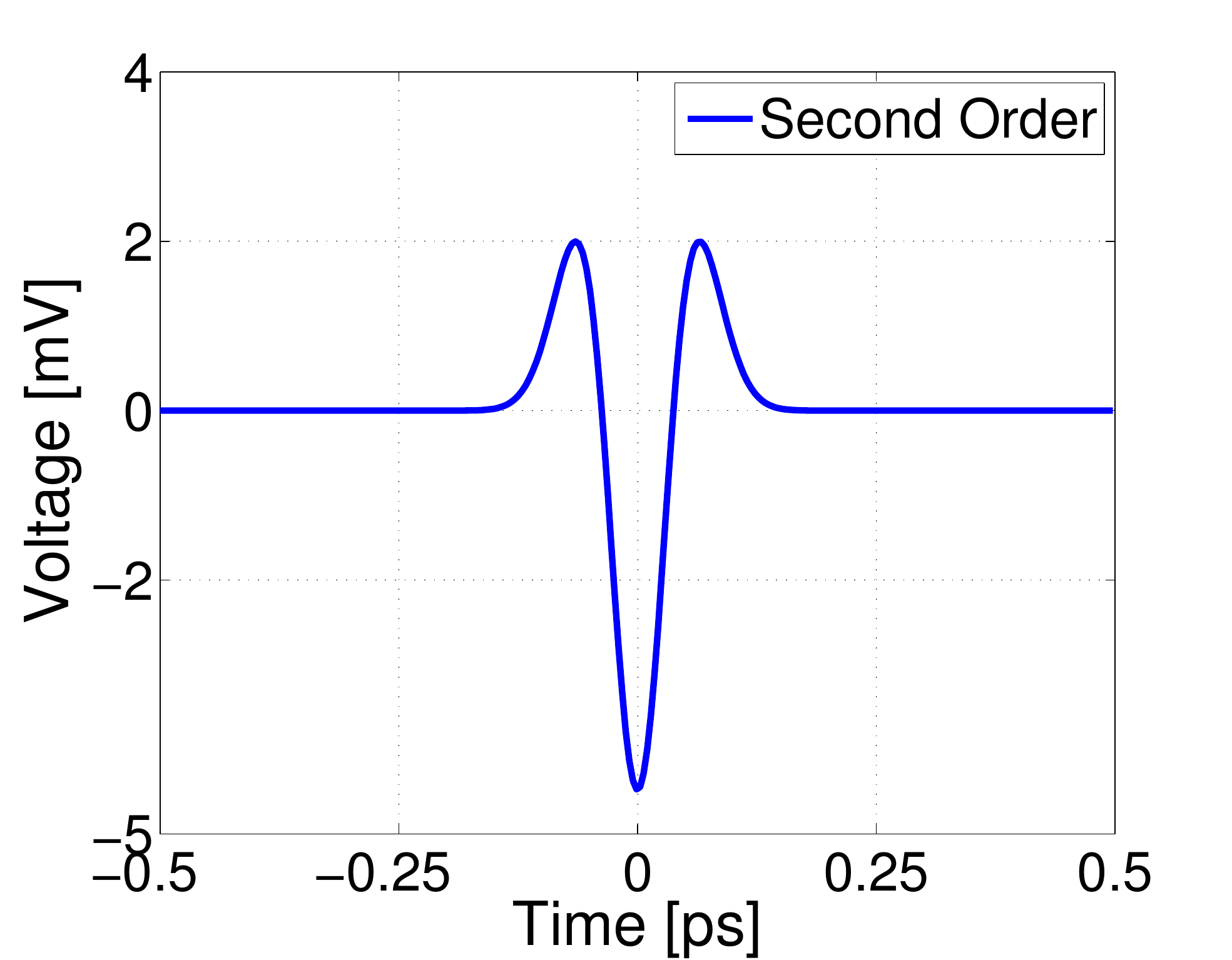}
			\label{fig:SECO}  }
		\subfigure[~]{
			\vspace{-4mm}
			\includegraphics[width=0.5\columnwidth, height = 3.25cm]{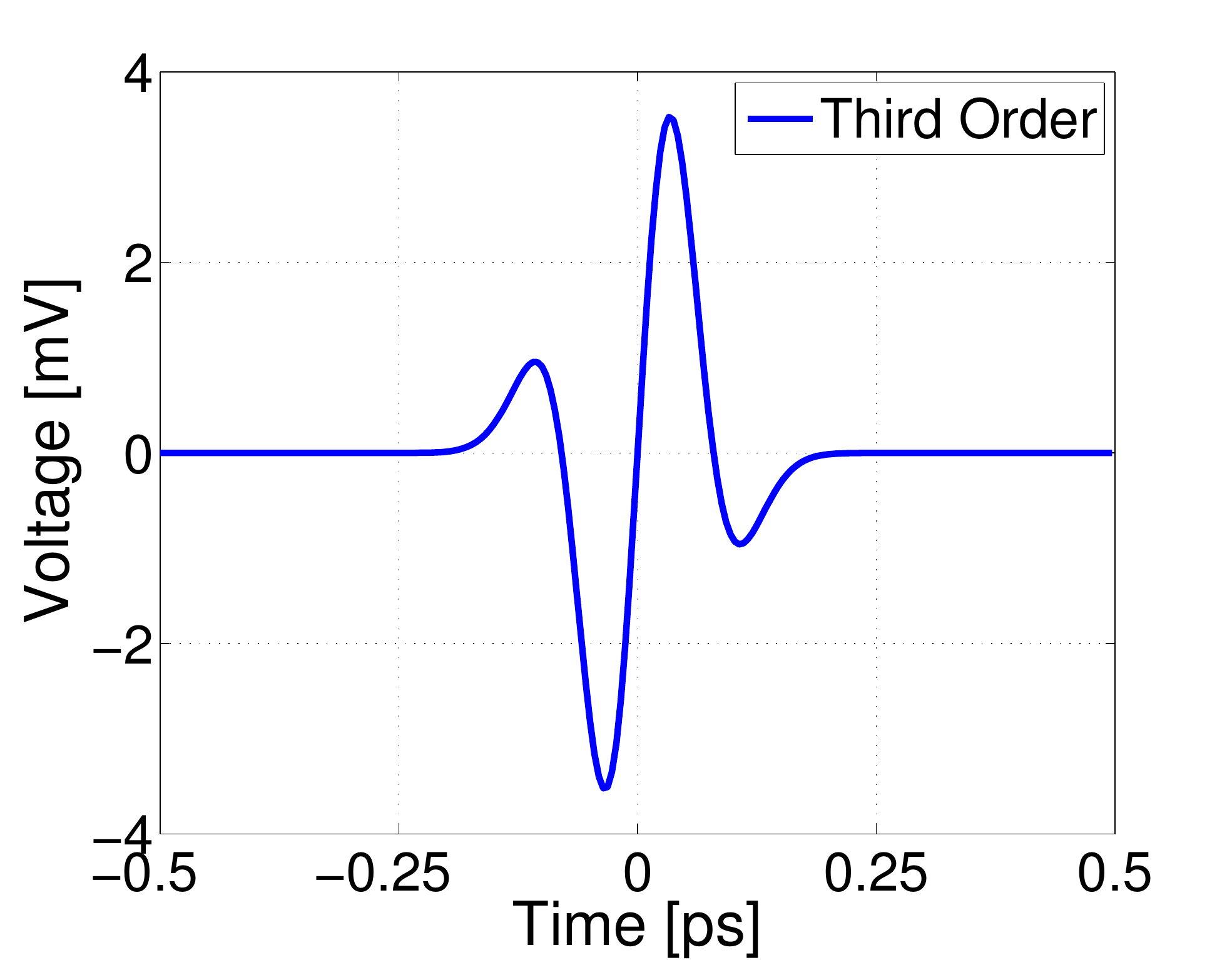}
			\label{fig:THRO}  }
		  \hspace{-5mm}
		\subfigure[~]{
			\vspace{-4mm}
			\includegraphics[width=0.5\columnwidth, height = 3.25cm]{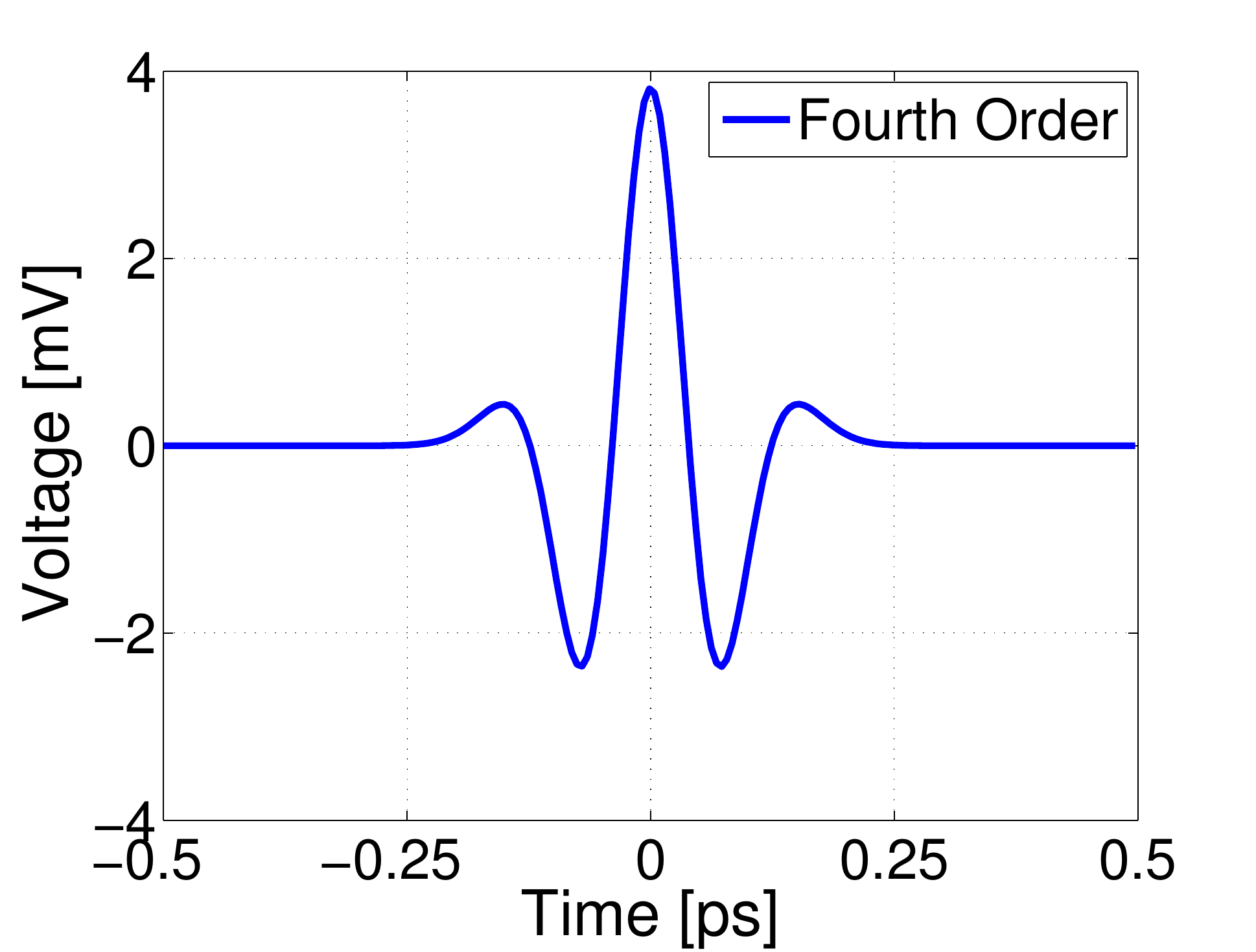}
			\label{fig:FOURO} }
		\subfigure[~]{
			\vspace{-4mm}
			\includegraphics[width=0.5\columnwidth, height = 3.25cm]{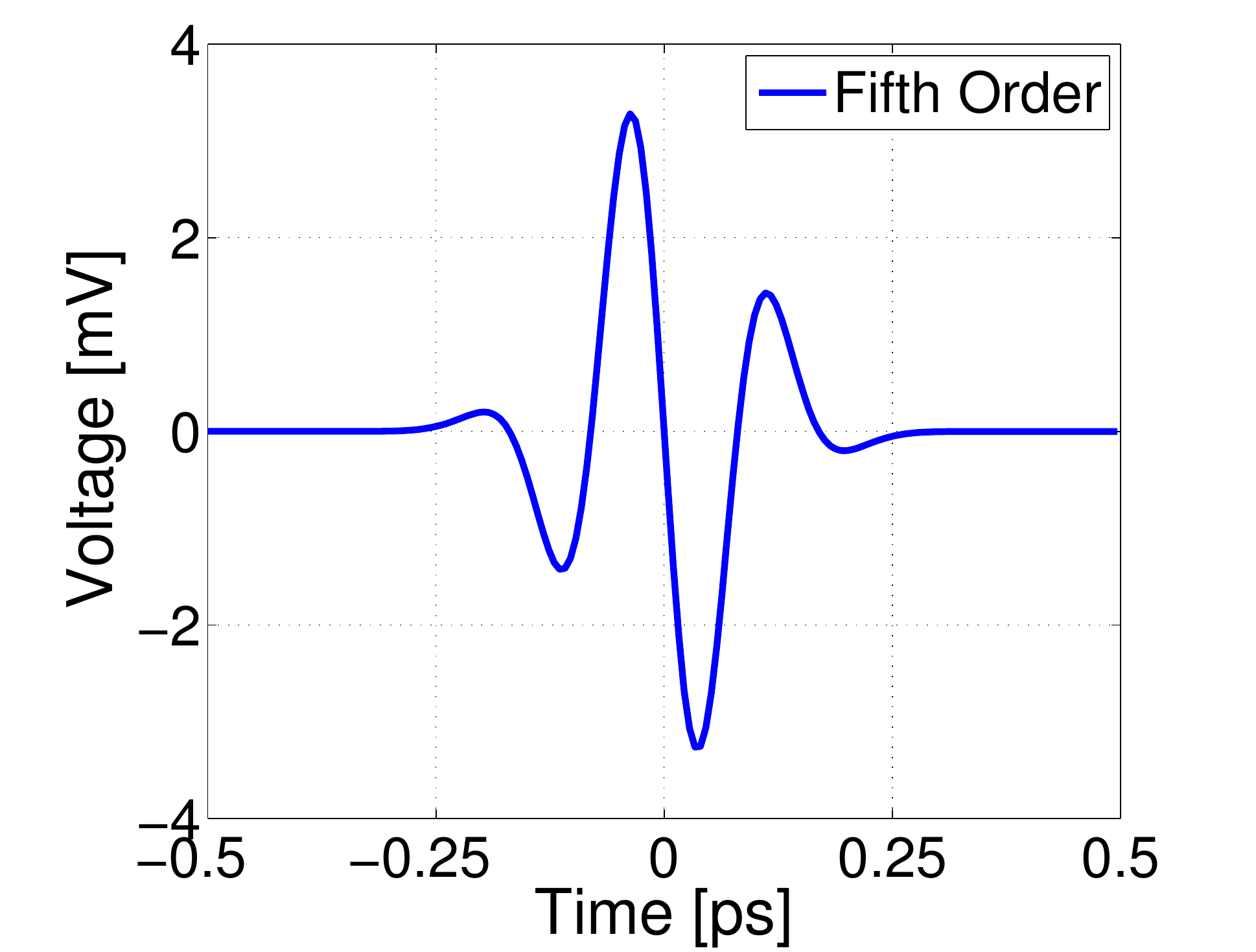}
			\label{fig:FIFO}  }
		\hspace{-5mm}
		\subfigure[~]{
			\includegraphics[width=0.5\columnwidth, height = 3.25cm]{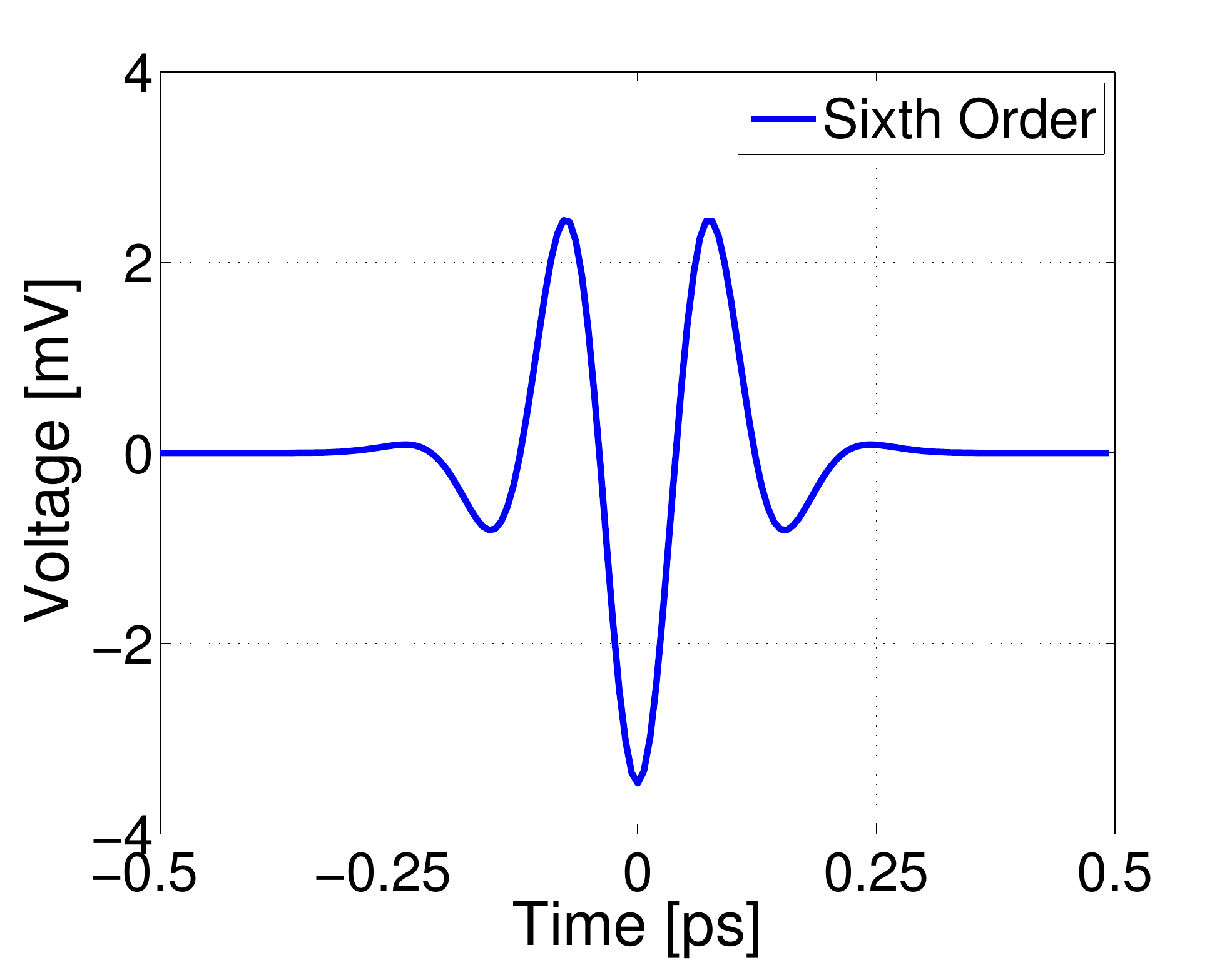}
			\label{fig:SIXO} }
		\vspace{-5mm}
		\caption{ Higher order time derivative Gaussian pulses of 1 aJ energy with 6 THz center frequency.}
		\label{fig:FIRPL}
		\vspace{-3mm}
	\end{figure}
	\begin{figure}[t]
		\centering
		\includegraphics[width=0.8\columnwidth, height =3.5cm]{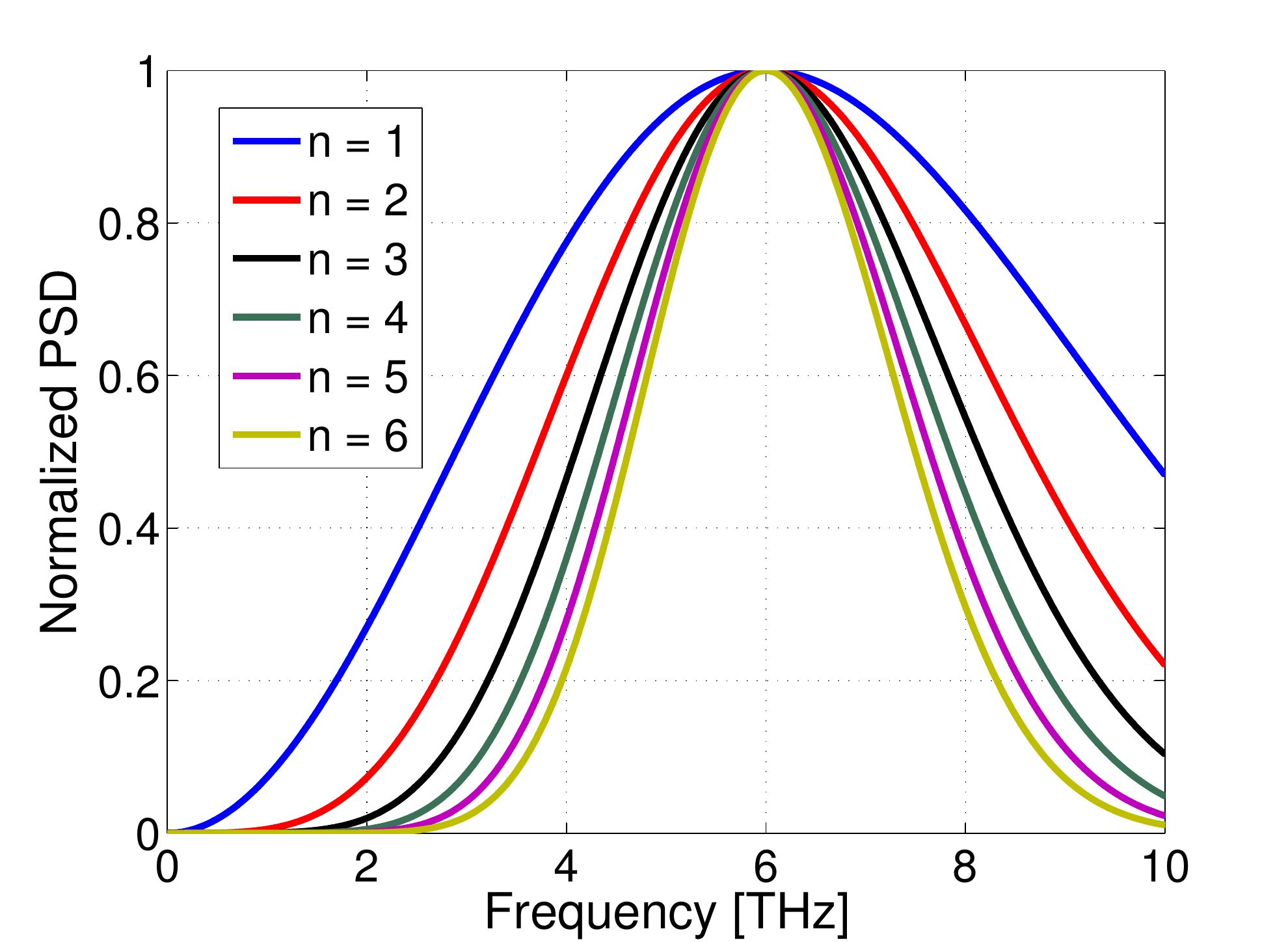}
		\vspace{-3mm}
		\caption{Normalized PSD of higher order Gaussian pulses with center frequency 6 THz.}
		\label{fig:PU_PSD}
		\vspace{-2mm}
	\end{figure}
	The number antenna elements $N$ in ULA is selected as 8 and distance between consecutive antenna elements is half the wavelength $\lambda_{min}$ of frequency 10 THz, that is $d_{s} = \SI{15}{\micro\meter}$ to avoid spatial aliasing. Since the DOA estimation accuracy decreases with smaller antenna spacing in ULA, terahertz band is considered from 1 THz onwards in the simulation. Thus, the maximum propagation time $\Delta T_{max}$ across the ULA is  0.35 ps. For ULA with aperture $D = \left( N-1\right) d_{s}$, the nanodevice source is assumed to be at a distance greater than $\left( 2*D^{2}\right) \textfractionsolidus\lambda_{min}$, the far-field region of ULA. The DOA of nanosensor device is considered as $10.25^{\circ}$. Since $\Delta T\gg \Delta T_{max}$ \cite{OPVT}, the observation time interval $\Delta T$ is assumed as 10 ps. Hence from \eqref{eq:timbw}, the number of frequency bins $L = 91$ for the terahertz bandwidth $B = 9\:THz$. The total number of frequency snapshots $K$ available is assumed as 50 and $N_{run}$ is selected as 100.\\
	In NWSN, nanosensor devices communicate by transmitting higher order time derivative Gaussian pulses. The time domain representation Gaussian pulses and its corresponding normalized PSD from time derivative order $n = 1 \;\text{to}\; 6$ is shown in Fig. \ref{fig:FIRPL} and Fig.\ref{fig:PU_PSD} respectively. The center frequency and energy of these higher order Gaussian pulses is $6 \:\text{THz} $ and 1 aJ, respectively. In table \ref{table:HPF}, the half power frequencies $f_{l}$ and $f_{h}$ in THz and half power bandwidth $B_{3\: dB}$ in THz is summarized for different higher order Gaussian pulses at different center frequencies $f_{c}$ in THz , $T_{p}$ represents duration of pulse in ps and $n$ represents the order of Gaussian pulse.
	%
	\begin{table*}[]
		\vspace{-5mm}
		\centering
		\caption{Half power bandwidth of higher order Gaussian pulses at different center frequencies}
		\vspace{-4.5mm}
		\resizebox{\textwidth}{!}{
			\begin{tabular}{|c|c|c|c|c|c|c|c|c|c|c|c|c|c|c|c|c|c|c|c|c|c|c|c|c|}
				\hline
				\multicolumn{1}{|l|}{} & \multicolumn{4}{c|}{$n$  = 1} & \multicolumn{4}{c|}{$n$  = 2} & \multicolumn{4}{c|}{$n$  = 3} & \multicolumn{4}{c|}{$n$  = 4} & \multicolumn{4}{c|}{$n$  = 5} & \multicolumn{4}{c|}{$n$  = 6} \\ \hline
				$f_{c}$ & $T_{p}$ & $f_{l}$ & $f_{h}$ & $B_{3\:dB}$ & $T_{p}$ & $f_{l}$ & $f_{h}$ & $B_{3\:dB}$ & $T_{p}$ & $f_{l}$ & $f_{h}$ & $B_{3\:dB}$ & $T_{p}$ & $f_{l}$& $f_{h}$ & $B_{3\:dB}$ &$T_{p}$ & $f_{l}$ & $f_{h}$ & $B_{3\:dB}$ & $T_{p}$ & $f_{l}$ & $f_{h}$ & $B_{3\:dB}$ \\ \hline
				2 & 0.79 & 0.96 & 3.27 & 3.27 & 1.12 & 1.23 & 2.88 & 1.64 & 1.37 & 1.36 & 2.71 & 1.35 & 1.59 & 1.44 & 2.61 & 1.71 & 1.77 & 1.49 & 2.54 & 1.04 & 1.94 & 1.54 & 2.49 & 0.95 \\ \hline
				3 & 0.53 & 1.44 & 4.90 & 3.46 &0.75 & 1.85 & 4.32 & 2.47 & 0.91 & 2.04 & 4.07 & 2.02 & 1.06 & 2.16 & 3.92 & 1.75 & 1.18 & 2.24 & 3.82 & 1.57 & 1.29 & 2.31 & 3.74 & 1.43 \\ \hline
				4 & 0.39 & 1.92 & 6.54 & 4.61 & 0.56 & 2.46 & 5.76 & 3.29 & 0.68 & 2.72 & 5.42 & 2.70 & 0.79 & 2.88 & 5.22 & 2.34 & 0.88 & 2.99 & 5.09 & 2.09 & 0.97 & 3.08 & 4.99 & 1.91 \\ \hline
				5 & 0.31 & 2.40 & 8.18 & 5.77 & 0.45 & 3.08 & 7.20 & 4.12 & 0.55 & 3.40 & 6.78 & 3.37 & 0.63 & 3.60 & 6.53 & 2.92 & 0.71 & 3.74 & 6.36 & 2.62 & 0.77 & 3.85 & 6.24 & 2.39 \\ \hline
				6 & 0.26 & 2.88 & 9.81 & 6.92 & 0.37 & 3.70 & 8.64 & 4.94 & 0.45& 4.09 & 8.14 & 4.05 & 0.53 & 4.32 & 7.84 & 3.51 & 0.50 & 4.49 & 7.64 & 3.14 & 0.64& 4.62 & 7.49 & 2.87 \\ \hline
		\end{tabular}}
		\label{table:HPF}
		\vspace{-4mm}
	\end{table*}
	\vspace{-6mm}
	\subsection{Impact of order and center frequency}
\begin{figure}[H]
		\vspace{-4mm}
		\centering
		\subfigure[~]{
			\includegraphics[width=4.25cm, height = 3.25cm]{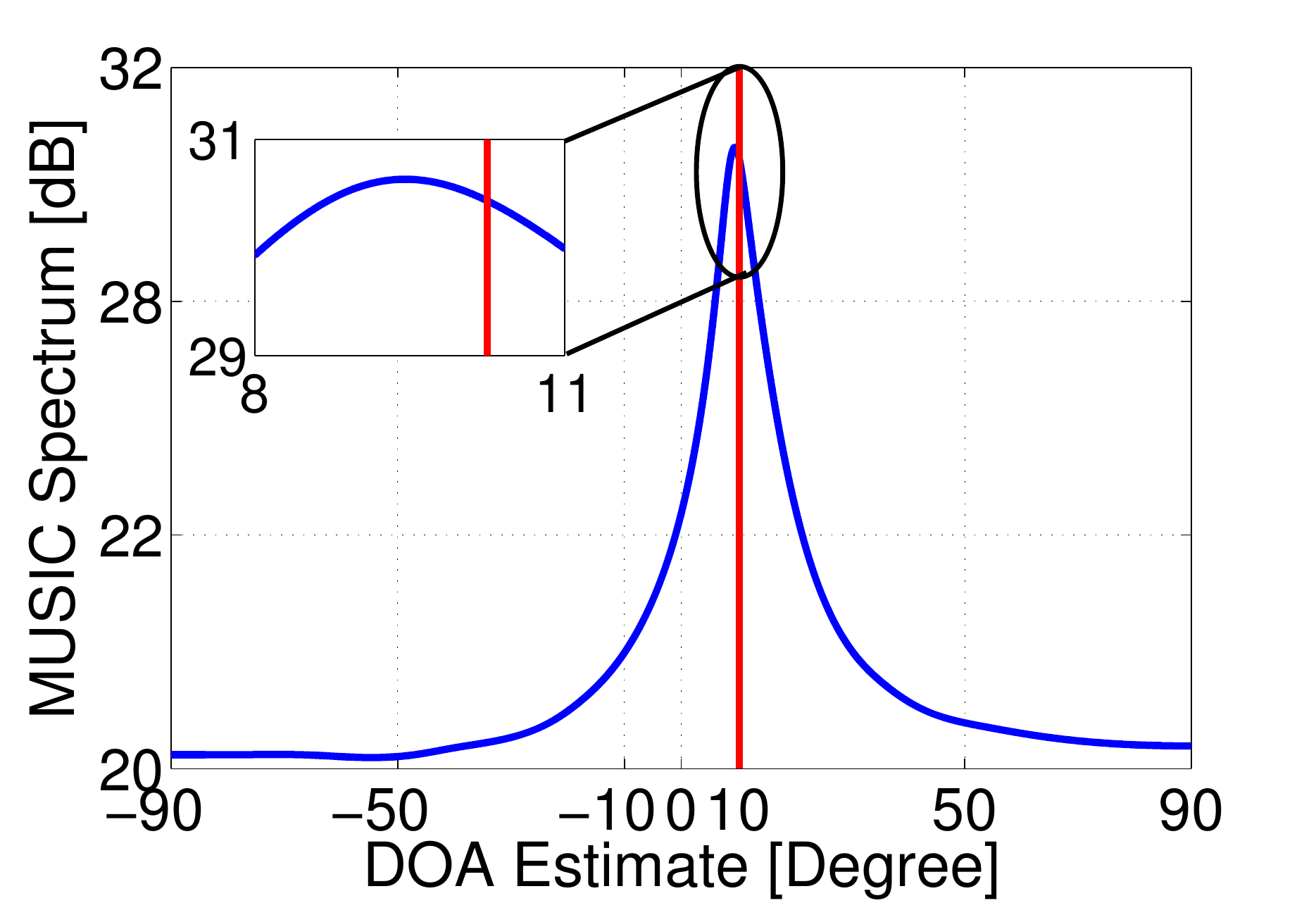}
			\label{fig:MUS_1_2T}
		}
	 \hspace{-5.5mm}
		\subfigure[~]{
			\vspace{2mm}
			\includegraphics[width=4.25cm, height = 3.25cm]{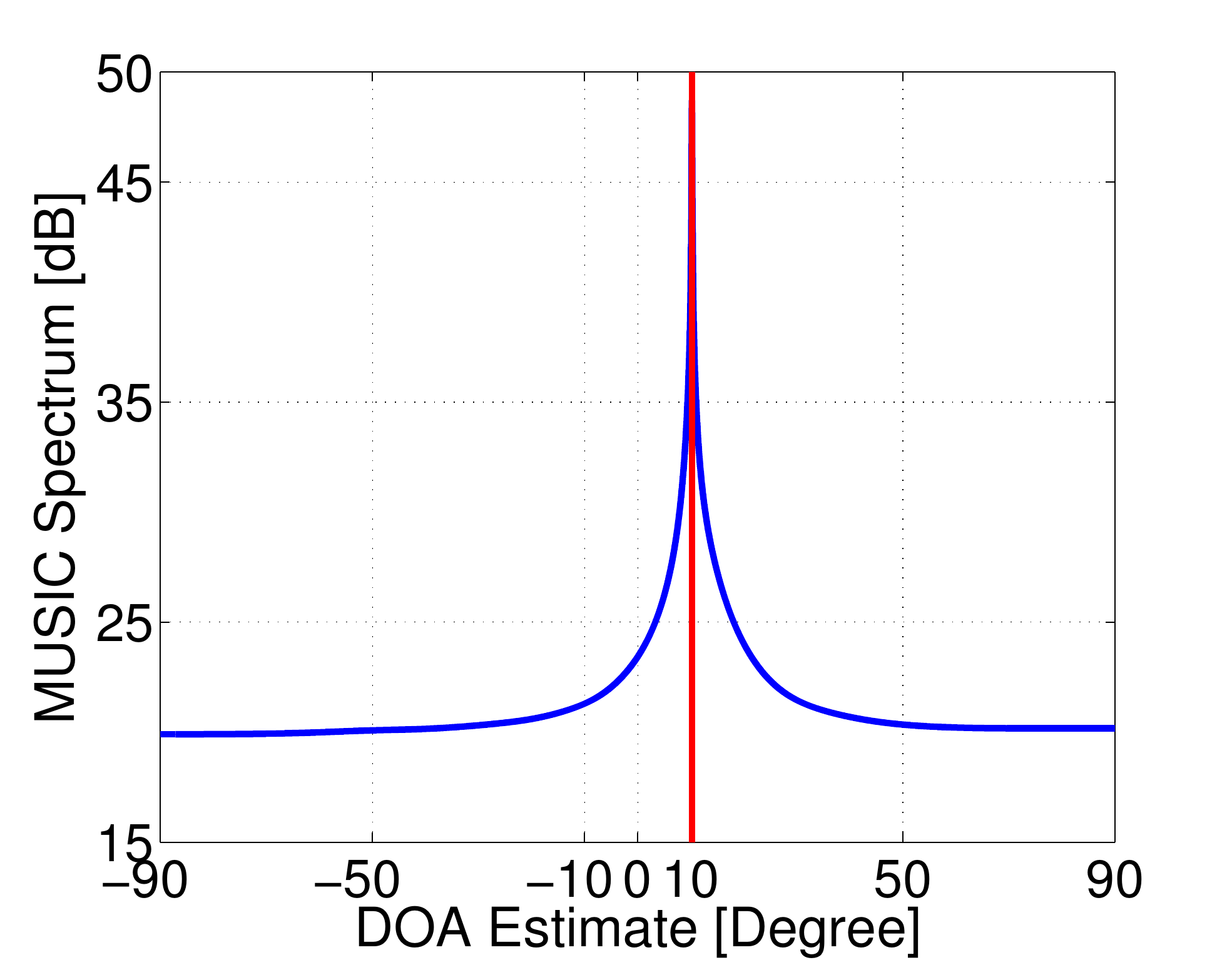}
			\label{fig:MUS_1_6T}  }
		\vspace{-6mm}
		\caption{IMUSIC Spectrum for first order Gaussian for two different center frequencies. The pulse energy is 0.01 aJ and the transmission distance is 0.5 m    \subref{fig:MUS_1_2T} Center frequency is 2 THz.   
			\subref{fig:MUS_1_6T} Center frequency is 6 THz.}
		\label{fig:MUS_1}
		\vspace{-4.5mm}
	\end{figure}

\begin{figure}[H]
		\centering
		\vspace{-4mm}
		\subfigure[~]{
			\includegraphics[width=4.25cm, height = 3.25cm]{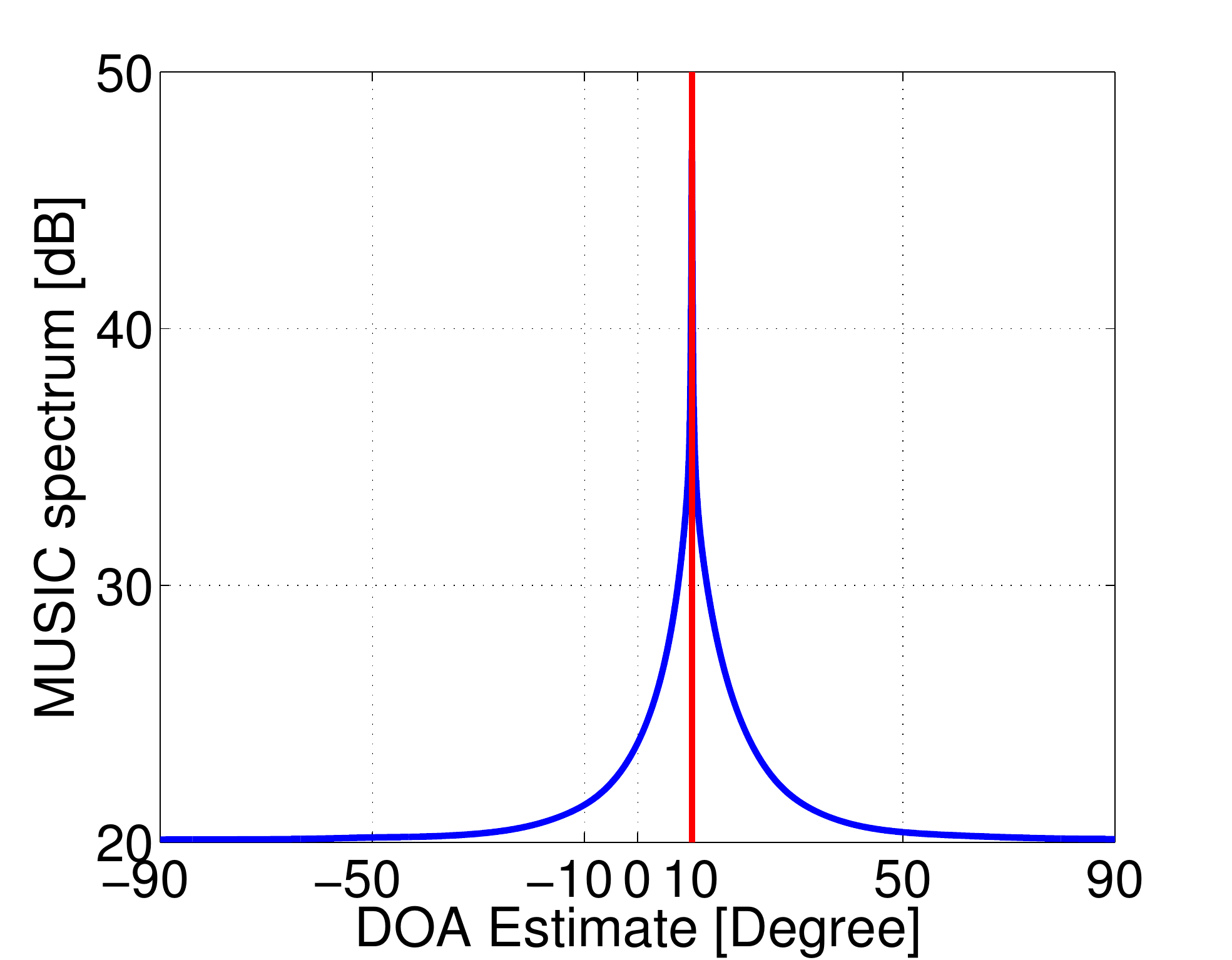}
			\label{fig:MUS_1_4T}
		}
	\hspace{-5.5mm}
		\vspace{-3mm}
		\subfigure[~]{
			\includegraphics[width=4.25cm, height = 3.25cm]{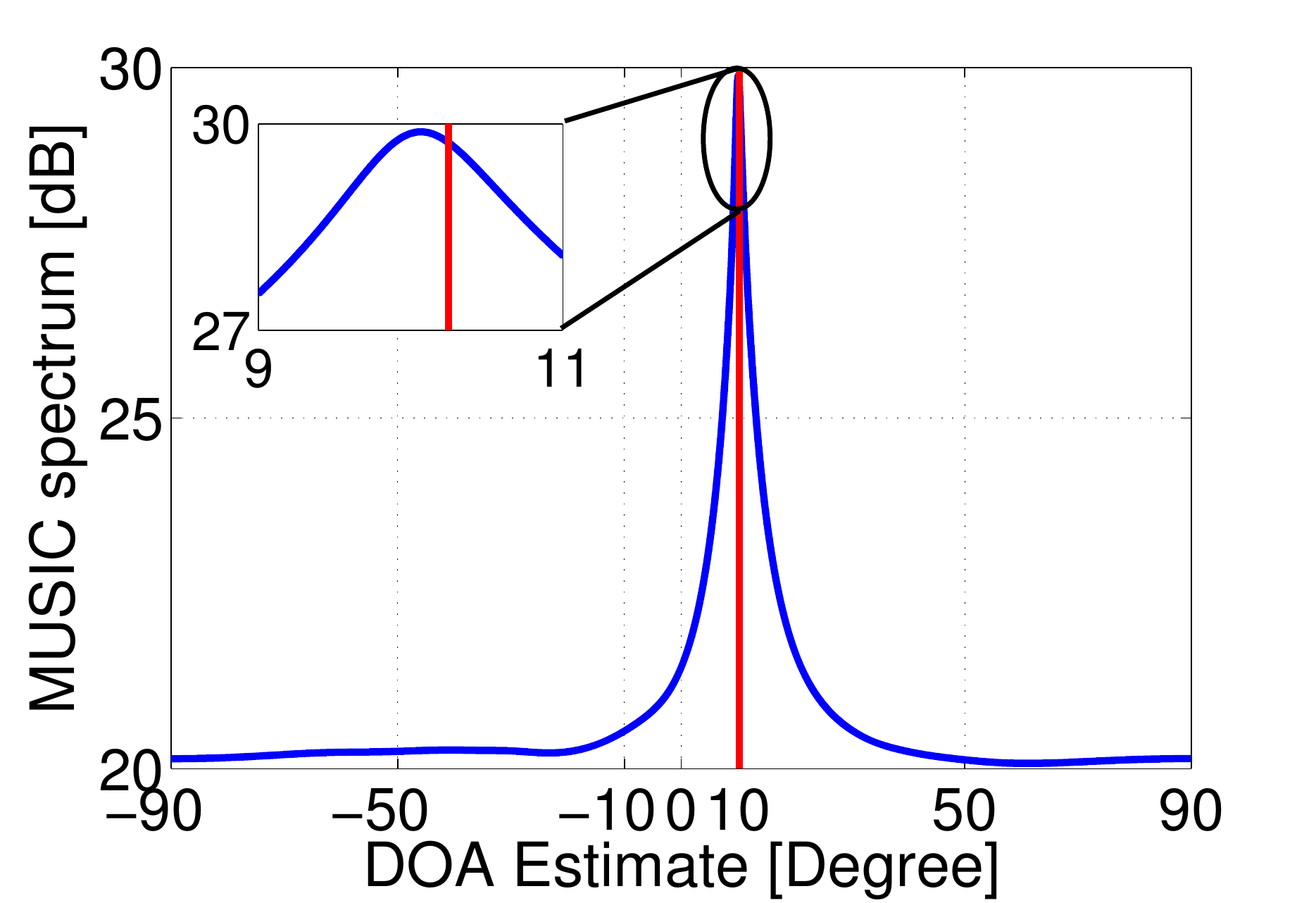}
			\label{fig:MUS_6_4T}  }
		\vspace{-2.5mm}
		\caption{ MUSIC Spectrum for nanosensor device transmitting first and sixth Gaussian pulse for center frequency 4 THz. The energy of both the Gaussian pulse are 0.01 aJ and the transmission distance is 0.5 m    \subref{fig:MUS_1_4T} Firth order Gaussian pulse.   
			\subref{fig:MUS_6_4T} Sixth order Gaussian pulse.}
		\label{fig:MUS_4T}
		\vspace{-2.5mm}
	\end{figure}

	To investigate the impact of order and frequency on DOA estimation error, we first examine their impact on IMUSIC spectrum. The impact of different center frequencies, 2 THz vs. 6 THz, for the same order (first order) is shown in Figs. \ref{fig:MUS_1_2T} and \ref{fig:MUS_1_6T}, respectively.  Similarly, the impact of different orders, first vs. sixth, for the same center frequency (4 THz) is shown in Figs. \ref{fig:MUS_1_4T} and \ref{fig:MUS_6_4T}, respectively. 

\begin{figure}
		\centering
		\includegraphics[width=\columnwidth, height =3.7cm]{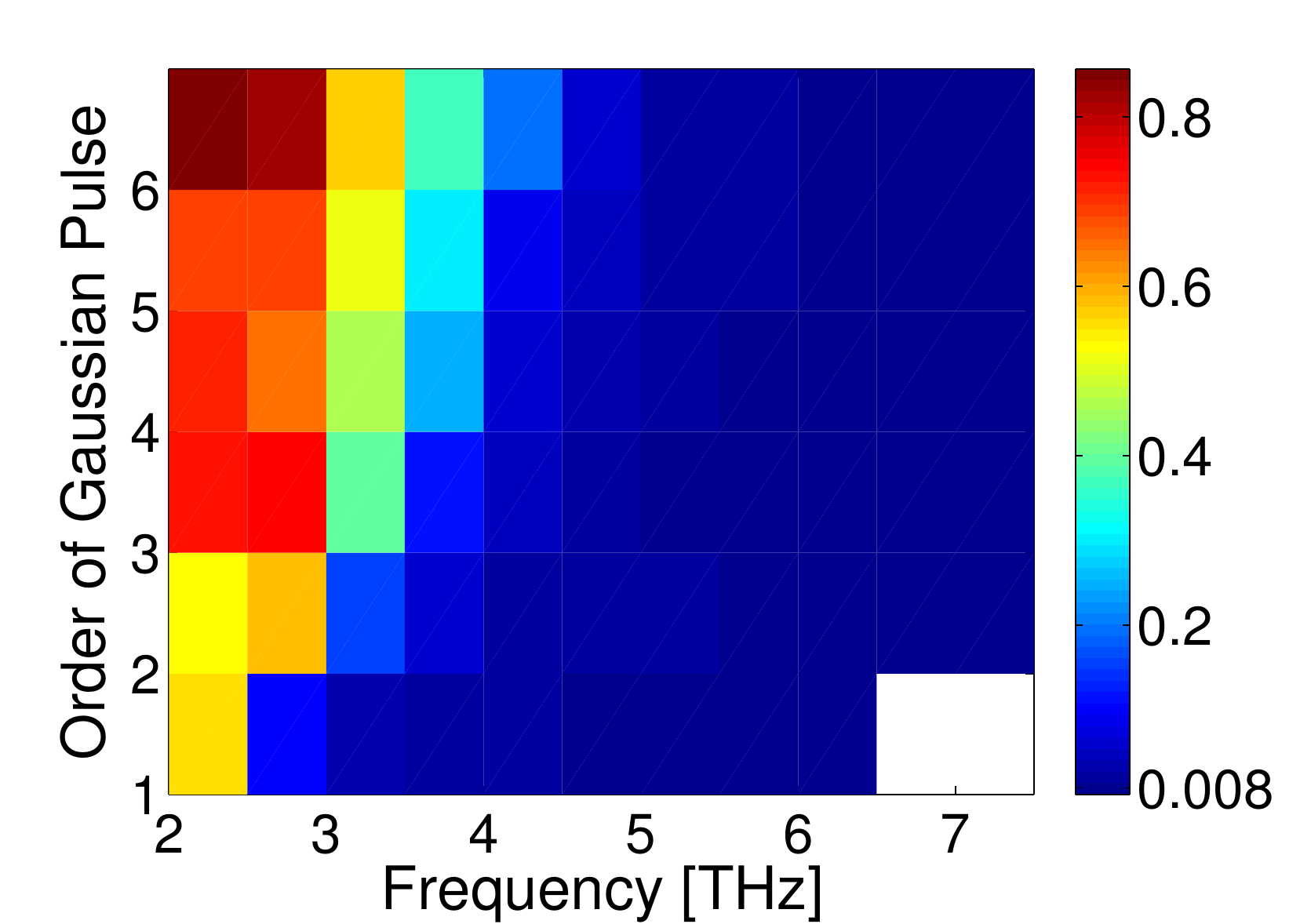}
		\vspace{-3mm}
		\caption{DOA estimation accuracy as a function of frequency and derivative order of 0.01 aJ Gaussian pulses transmitted from a distance of 0.5 m}
		\vspace{-2mm}
		\label{fig:TEMP_GRAP}
	\end{figure}

	We can see that the peak of IMUSIC spectrum is closer to the DOA value (10.25 degree) for the higher frequency (6 THz in Fig. \ref{fig:MUS_1_6T}) and lower order (first order in Fig. \ref{fig:MUS_1_4T}). A possible explanation for this outcome is because the higher frequencies and lower orders lead to higher 3 dB bandwidth and shorter $T_{p}$ (see Table \ref{table:HPF}). Note that shorter $T_{p}$ means that the pulse can be transmitted with higher power for the same amount of energy. 
	Next, we focus on RMSE values to quantify DOA estimation as a function of center frequencies and derivative orders of Gaussian pulses. The temperature graph in Fig.\ref{fig:TEMP_GRAP} plots these RMSE values for pulse energy of 0.01 aJ and distance of 0.5 m (best viewed in color).We can clearly observe that RMSE decreases with lower orders and higher frequencies. This suggests that for NWSNs, the best performance for DOA can be achieved with the first order Gaussian pulses emitting their peak energies at 6 THz center frequency. Thus, in table \ref{my-label}, we derive the RMSE values for different distances for first order and 6 THz center frequency and 1 aJ energy. \textit{We can see that IMUSIC with $K = 1$ frequency snapshot can achieve DOA estimation accuracy with RMSE less than one hundredth of a degree from a distance of 6 meter for pulse energy as little as 1 atto Joule .} This is a remarkable performance for IMUSIC, which can enable many applications in NWSNs.
	\begin{table}[]
		\centering
		\vspace{-1mm}
		\caption{RMSE values at different distances for first order Gaussian pulse with energy 1 aJ and center frequency 6 THz with $K = 1$ frequency snapshots.}
		\vspace{-4.5mm}
		\label{my-label}
		\resizebox{\columnwidth}{0.08\columnwidth}{
			\begin{tabular}{|c|c|c|c|}
				\hline
				\textbf{Distance {[}m{]}} & \textbf{RMSE} & \textbf{Distance {[}m{]}} & \textbf{RMSE} \\ \hline
				0.01	&0.010105939      &               3            &0.275689935\\ \hline
				0.1	&     0.023144645       &                    5       &0.500921665 \\ \hline
				1	&   0.100274849              &                   6        &0.600167844 \\ \hline
		\end{tabular}}
		\vspace{-2mm}
	\end{table}
\vspace{-1mm}
	\subsection{Impact of Distance}
\begin{figure*}[t]
		\vspace{-2mm}
		\centering
		\subfigure[~]{
			\vspace{-1mm}
			\includegraphics[width=4.4cm, height = 3.25cm]{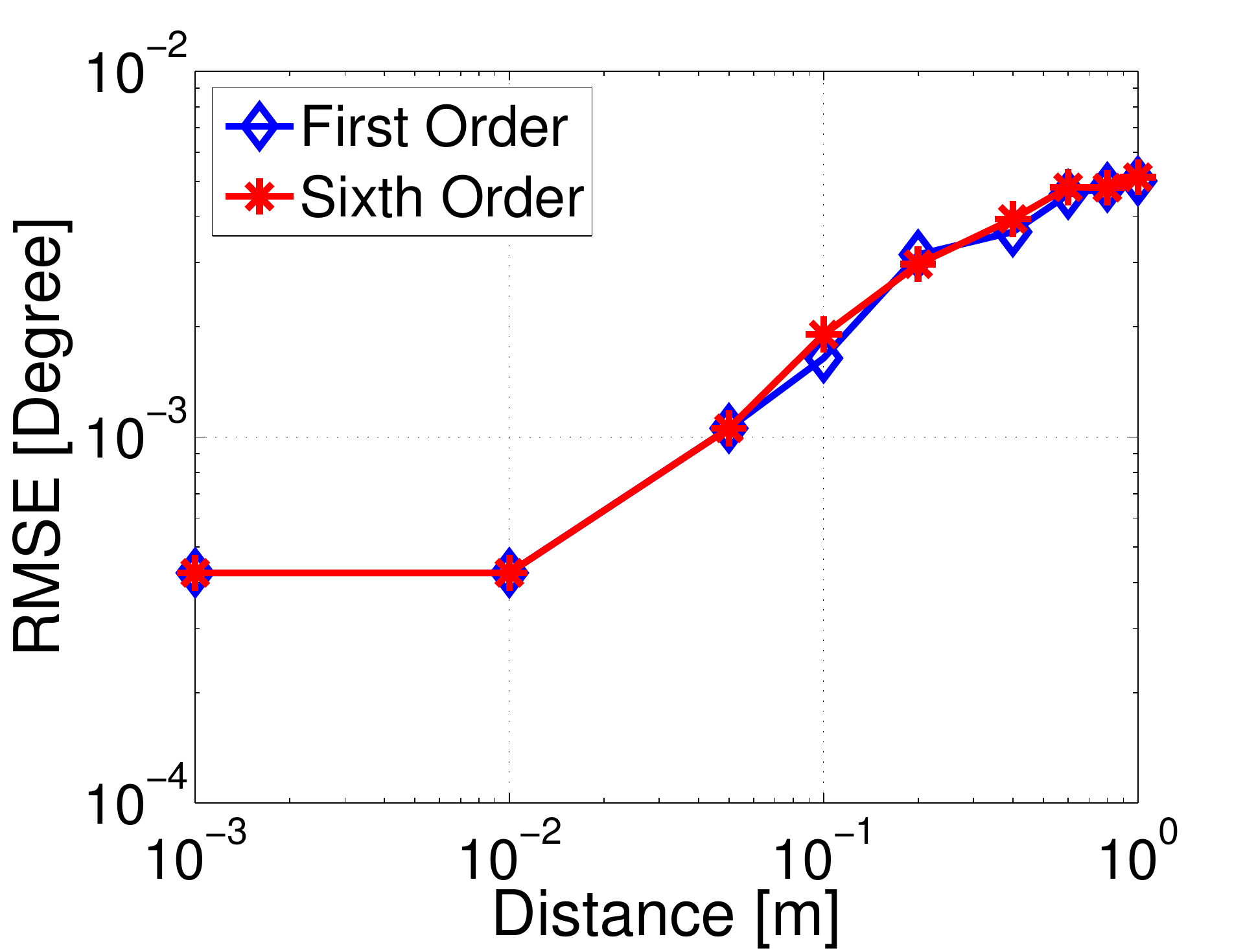}
			\label{fig:FX6_100}
		}
	   \hspace{-4mm}
		\vspace{-1mm}
		\subfigure[~]{
			\vspace{-1mm}
			\includegraphics[width=4.4cm, height = 3.25cm]{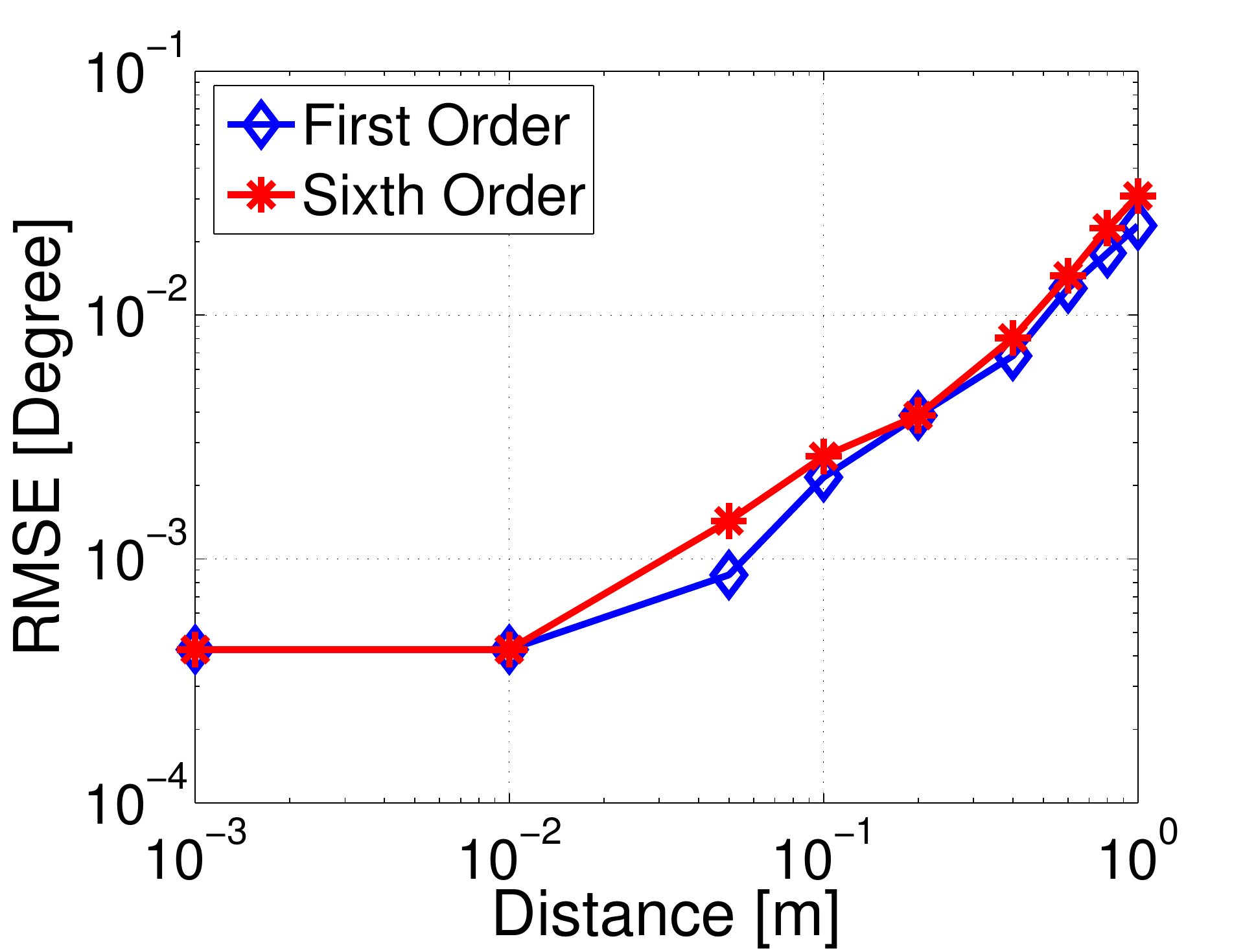}
			\label{fig:FS6_001}  }
		\hspace{-4mm}
		\vspace{-1mm}
		\subfigure[~]{
			\includegraphics[width=4.4cm, height = 3.25cm]{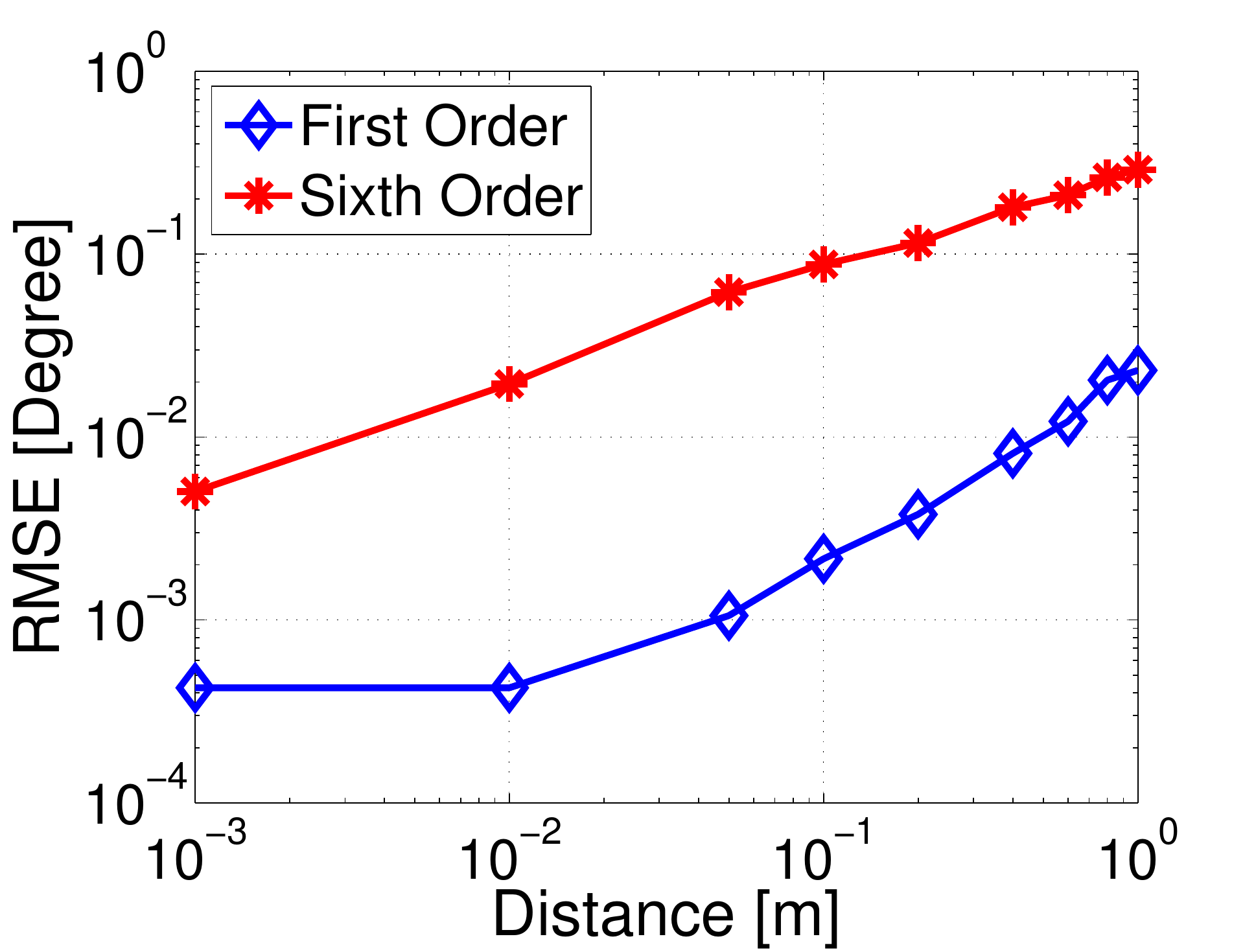}
			\label{fig:FS2_100}  }
		\hspace{-4mm}
		\vspace{-1mm}
		\subfigure[~]{
			\vspace{-1mm}
			\includegraphics[width=4.4cm, height = 3.25cm]{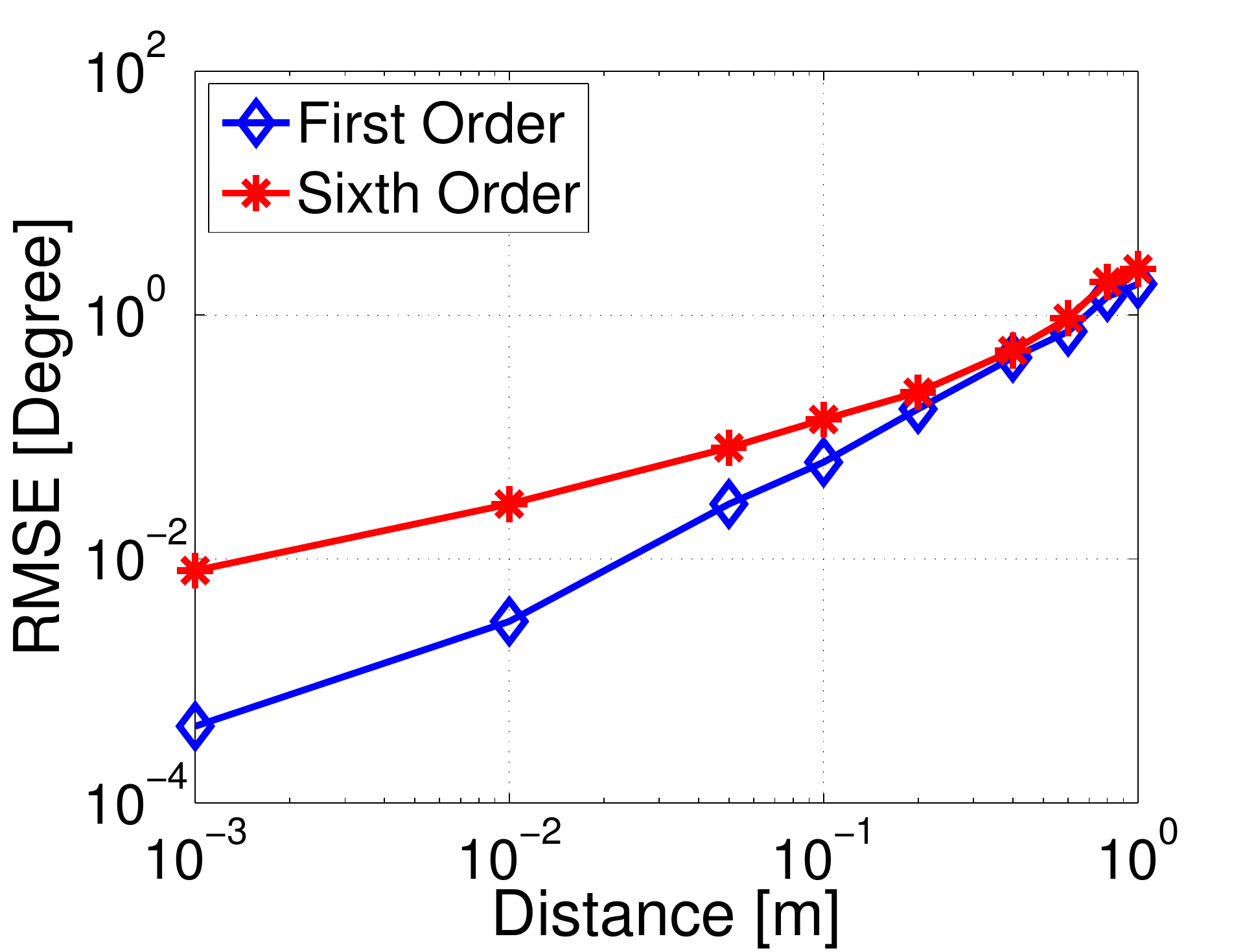} 
			\label{fig:FS2_001} }
		\vspace{-1.0mm}
		\caption{RMSE versus Distance for first and sixth order Gaussian pulses for different energy values\subref{fig:FX6_100} First and Sixth order Gaussian pulse with energy 100 aJ and center frequency 6 THz.   
			\subref{fig:FS6_001} First and Sixth order Gaussian pulse with energy 0.01 aJ and center frequency 6 THz.
			\subref{fig:FS2_100} First and Sixth order Gaussian pulse with energy 100 aJ and center frequency 2 THz.
			\subref{fig:FS2_001} First and Sixth order Gaussian pulse with energy 0.01 aJ and center frequency 2 THz.}
		\vspace{-2mm}
		\label{fig:RMSE_V_DIS}
	\end{figure*}

	Fig. \ref{fig:RMSE_V_DIS} shows the DOA estimation performance as a function of distance between the emitter and the ULA. We make the following interesting observations: (1) For high frequency (6 THz), distance has no significant impact for up to 1 cm, but RMSE starts to increase rapidly after that (Figs.\ref{fig:FX6_100} and \ref{fig:FS6_001}). A possible explanation for this outcome is due to negligible effect of molecular absorption noise for shorter transmission distances. Further, at this frequency, there is no significant difference between derivative orders of the Gaussian pulse, and (2)  For low frequency (2 THz), RMSE increases for increasing distance even below 1 cm except for first order with high energy (100 aJ) as shown in Figs.\ref{fig:FS2_100} and \ref{fig:FS2_001}. At this frequency, first order clearly outperforms 6th order, but for low energy (0.01 aJ in \ref{fig:FS2_001}), RMSE for first order increases at a faster rate than that of 6th order. This rapid increase in RMSE for increasing distance is due to the increase in the effect of molecular absorption noise and terahertz path loss for large distances.
	%
	\vspace{-2mm}
	\vspace{-3mm}
	\subsection{Impact of Pulse Energy}
\begin{figure}[H]
		\centering
		\includegraphics[width=0.85\columnwidth, height =3.5cm]{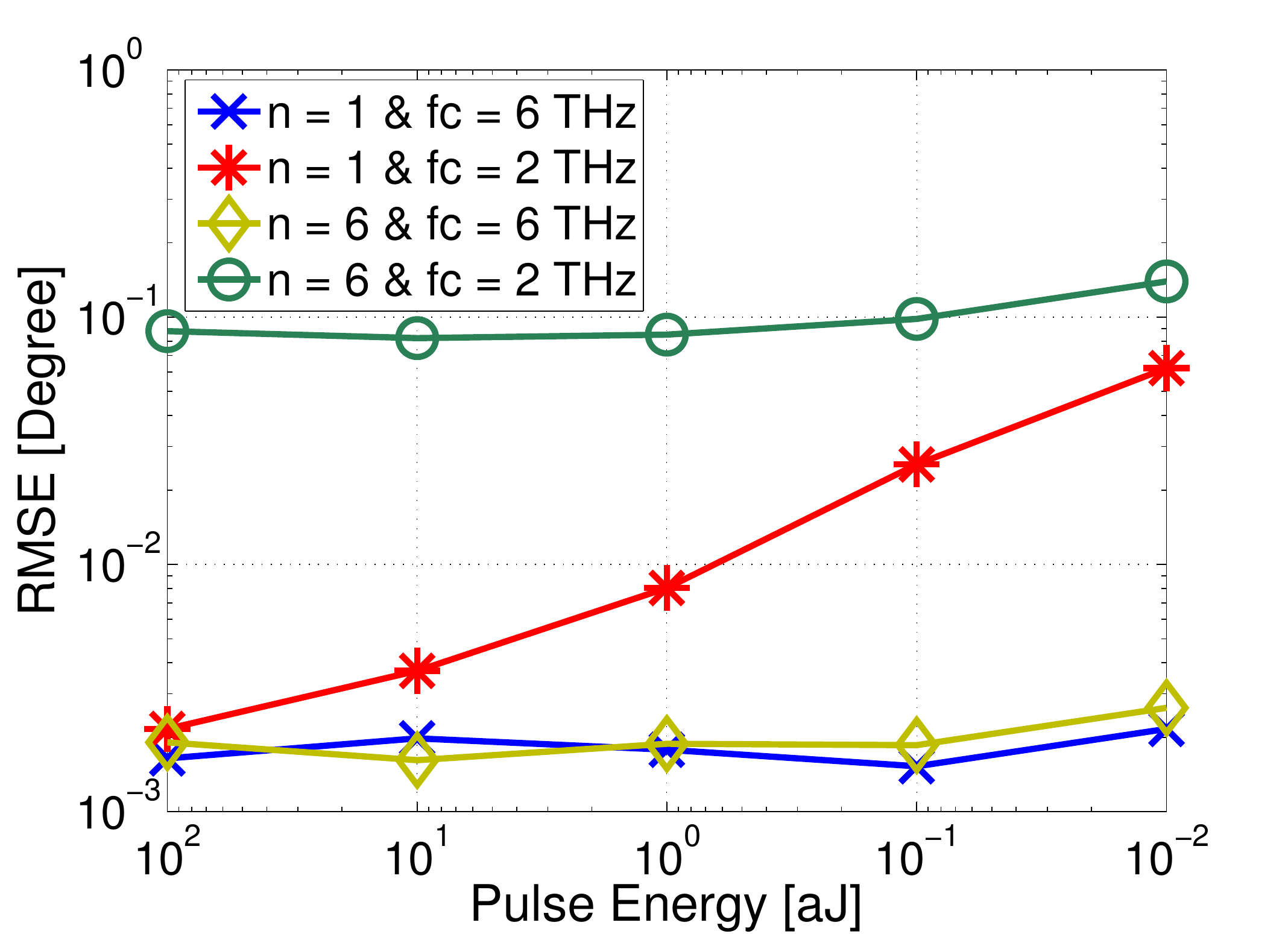}
		\vspace{-2mm}
		\caption{RMSE versus Pulse Energy for first and sixth order Gaussian pulse for two center frequencies 2 THz and 6 THz. Distance between ULA and nanosensor device is 0.1 m.}
		\label{fig:RMSE_ERG}
	\end{figure}

	The dependence of DOA estimation accuracy on pulse energy is investigated in Fig.  \ref{fig:RMSE_ERG}. We make the following interesting observations: (1) For high center frequency (6 THz), pulse energy has little effect on DOA performance (bottom two curves), (2) pulse energy has little impact on DOA estimate for high order (sixth order) and low center frequency of 2 THz (top curve), and (3) Pulse energy, however, has significant effect on DOA performance for low order (first order) and low frequency (2 THz) pulses. In this case, RMSE increases rapidly with decreasing pulse energy (see red curve).
	\vspace{-5mm}
	\subsection{Impact of frequency snapshots}
\begin{figure}[H]
		\centering
		\includegraphics[width=0.7\columnwidth, height = 2.75cm]{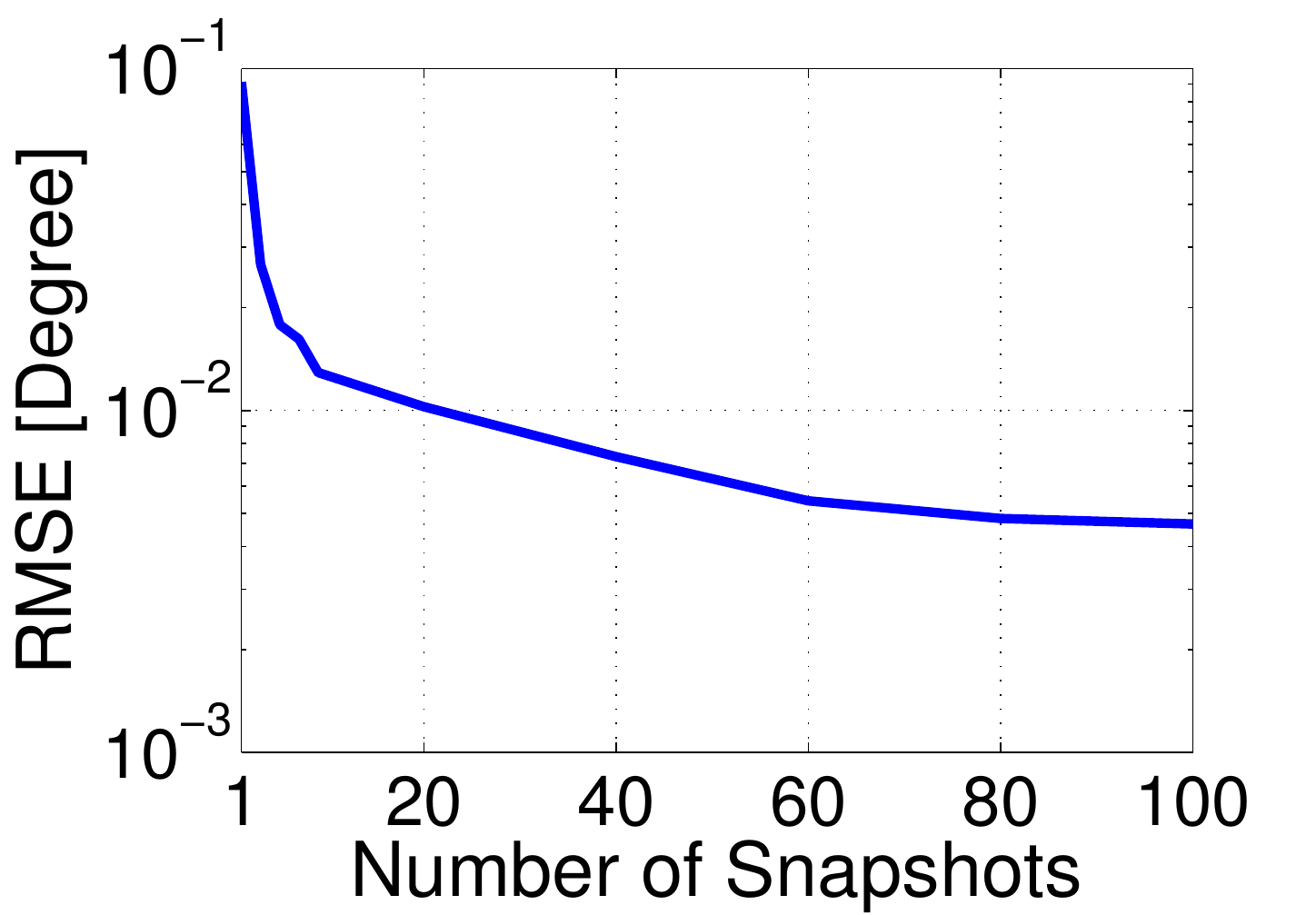}
		\vspace{-2.5mm}
		\caption{RMSE versus number of frequency snapshots. The energy of first order Gaussian pulses with center frequency 6 THz is 1 aJ and distance is assumed as 1 m.}
		\label{fig:snap}
		\vspace{-3.0mm}
	\end{figure}

	Finally, the performance of IMUSIC algorithm for different number of frequency snapshots is investigated in Fig.\ref{fig:snap} for first order Gaussian pulses with center frequency of 6 THz, pulse energy of 1 aJ, and distance of 1 m. It is observed from Fig.\ref{fig:snap} that RMSE decreases rapidly as the number of snapshots increase up to 50, after that the effect of increasing snapshots becomes less significant. This result provides important guidelines for selecting the number of snapshots for IMUSIC in NWSNs.  
	\section{Conclusion}
	\vspace{-1mm}
	Using wideband MUSIC algorithm, we have analyzed DOA estimation for terahertz NWSNs for higher order Gaussian pulses with varying center frequency, energy, transmission distance and frequency snapshot values. Our investigation shows that RMSE error can be reduced by selecting lower order and higher frequency pulses for transmissions. For first order Gaussian pulse emitting its peak energy at 6 THz, MUSIC can keep RMSE below 1 degree from a distance of 6 meter for pulse energy as little as 1 atto Joule with just a single snapshot. In future, DOA estimation performance can be investigated for multiple nanosensor devices using various other DOA estimation algorithm along with estimation of center frequency. 
	\vspace{-3mm}
	\bibliographystyle{ACM-Reference-Format}
	\bibliography{NANO_COM_BIBLIO_V1}
\end{document}